\documentclass[aps, prb, twocolumn, superscriptaddress, english]{revtex4-2}
\usepackage[english]{babel}
\usepackage{graphicx}
\usepackage{cancel}
\usepackage{enumerate}
\usepackage{hyperref}
\hypersetup{pdfencoding=auto,colorlinks=true,linkcolor=blue,citecolor=blue}
\usepackage{textcomp}
\usepackage{amsmath}
\usepackage{amssymb}
\usepackage{bbold}
\usepackage{pgf}
\usepackage{soul}
\usepackage{subfigure}
\usepackage{tikz}
\usepackage[normalem]{ulem}
\usepackage[dvipsnames]{xcolor}

\def\bra#1{\langle#1 |}
\def\ket#1{| #1\rangle}
\newcommand{\braket}[2]{\langle #1 | #2 \rangle}

\newcommand{\new}[1]{{\textcolor{black}{#1}}}

\newcommand{\se}{\kappa}
\newcommand{\me}{\upsilon}
\newcommand{\bracket}[2]{\langle#1|#2\rangle}
\newcommand{\taf}{\tilde{\tau}}

\graphicspath{{./}{./Figures/}}
\begin{document}
\title{When the center matters: color screening and gluelumps in dihedral lattice gauge theories}

 \author{Pavel P. Popov}
 \thanks{Equal contribution}
  \affiliation{ICFO - Institut de Ci\`encies Fot\`oniques, The Barcelona Institute of Science and Technology, 08860 Castelldefels (Barcelona), Spain}
 
\author{Edoardo Ballini}
 \thanks{Equal contribution}
 \affiliation{Pitaevskii BEC Center, CNR-INO and Dipartimento di Fisica, Universit\`a di Trento,  Via Sommarive 14, I-38123 Trento, Italy}
 \affiliation{INFN-TIFPA, Trento Institute for Fundamental Physics and Applications, Via Sommarive 14, I-38123 Povo, Trento, Italy}
 
\author{Alberto Bottarelli}
\affiliation{Pitaevskii BEC Center, CNR-INO and Dipartimento di Fisica, Universit\`a di Trento,  Via Sommarive 14, I-38123 Trento, Italy}
 \affiliation{INFN-TIFPA, Trento Institute for Fundamental Physics and Applications, Via Sommarive 14, I-38123 Povo, Trento, Italy}
 \affiliation{Honda Research Institute Europe GmbH, Carl-Legien-Str.\ 30, 63073 Offenbach, Germany}

\author{Michele Burrello}
\affiliation{Dipartimento di Fisica, Università di Pisa, and INFN, Sezione di Pisa, Largo Pontecorvo 3, I-56127 Pisa, Italy}

\author{Pietro Silvi}
\affiliation{Dipartimento di Fisica e Astronomia “G. Galilei” and Padua Quantum Technologies Research Center, Universit`a di Padova, I-35131 Padova, Italy}
\affiliation{Istituto Nazionale di Fisica Nucleare (INFN), Sezione di Padova, I-35131 Padova, Italy}

\author{Matteo M. Wauters}
  \affiliation{Pitaevskii BEC Center, CNR-INO and Dipartimento di Fisica, Universit\`a di Trento,  Via Sommarive 14, I-38123 Trento, Italy}
 \affiliation{INFN-TIFPA, Trento Institute for Fundamental Physics and Applications, Via Sommarive 14, I-38123 Povo, Trento, Italy}

\author{Philipp Hauke}
 \affiliation{Pitaevskii BEC Center, CNR-INO and Dipartimento di Fisica, Universit\`a di Trento,  Via Sommarive 14, I-38123 Trento, Italy}
 \affiliation{INFN-TIFPA, Trento Institute for Fundamental Physics and Applications, Via Sommarive 14, I-38123 Povo, Trento, Italy}

\begin{abstract}
  Confinement is one of the hallmarks of quantum chromodynamics (QCD), but its first-principles characterization remains elusive.
  We show that discrete non-Abelian lattice gauge theories (LGTs) with dihedral groups $D_N$ provide a minimal setup for characterizing the connection between confinement and the group center.
  When the latter is trivial (odd $N$), gluon clouds screen static charges, forming composite objects known as gluelumps. These are stable excitations of the theory, in contrast to their QCD analog. As a consequence, string breaking can occur through fluctuations of the electric field only, without pair nucleation from the vacuum.
  Our results showcase how the rich physics typically associated with QCD emerges in much simpler discrete non-Abelian LGTs, making them ideal settings to test this phenomenology both in numerical calculations and in near-term quantum devices. 
\end{abstract}

\maketitle
\emph{Introduction.---}
Deriving confinement of color-charged particles from first principles remains one of the central open challenges of QCD~\cite{Greensite2003}. Classical Monte Carlo simulations of lattice gauge theories (LGTs) have been extremely successful in addressing many nonperturbative phenomena~\cite{rebbi1983lattice,Kogut_RMP1983,Creutz_PhysRep1983,Brambilla2014,Aarts_2023}, yet they are hindered by fundamental limits when studying, e.g., real-time evolution and large-fermionic-density regimes. Hamiltonian-formulation LGTs, free of the sign problem, have therefore become a prime target for quantum simulation and tensor-network algorithms~\cite{Silvi2014,Zohar_PRD2015,dalmonte2016,lamm2019prd,banuls2020,Aidelsburger_LGT2021,Zohar_PhilTransA2021,Bauer_PRXQ2023, DiMeglio_PRXQ2024,Cataldi_PRR2024}, which have already probed equilibrium~\cite{Silvi_PRD2019,Rigobello_2023} and dynamical~\cite{Ciavarella_PRL2024,Hayata_PRD2025} properties of one-dimensional SU(3) LGTs, despite the challenges posed by the large number of degrees of freedom.

Hamiltonian simulations of LGTs require suitable approximations to encode the continuous degrees of freedom of the Lie group onto a finite Hilbert space. Typically, one resorts to a truncation in representation space~\cite{ Chandrasekharan_NucPhysB1997,Tagliacozzo_PRX2014,Zohar_NJP2016,Silvi2017,Felser_PRX2020,Popov_PRR2024} or in a dual basis~\cite{Mathur_PRD2015,Paulson_PRX2021,Haase2021,Fontana_PRX2025,Mirandariaza_2025}. 
Complementary methods are based on the quantum simulations of models with discrete gauge symmetries to approximate the related continuous\new{-group} parent theories~\cite{Notarnicola_JPA2015,Ercolessi2018,Gustafson_PRD2022,GonzalezCuadra_PRL2022,Mariani_PRD2023, Gustafson_PRD2024, Gustafson_PRD2024a, calliari_2025, perez2025primitivequantumgatessu3}. These models share many features with topologically ordered 2D systems~\cite{kitaev2002topological,Semeghini_Sci2021,Satzinger_Science2021,Andersen_Nature2023,Xu_2023,Xu_NatPhys2024,Iqbal2024}, 
and were instrumental to achieve the first quantum simulations of 2D dynamical string breaking phenomena~\cite{Cochran2025,Gonzalezcuadra_2025,Cobos2025}.
Overall, LGTs are identified as a testbed for possible quantum advantage and enjoy substantial interest beyond the high-energy physics community~\cite{DiMeglio_PRXQ2024}.

Dihedral groups $D_N$ are among the simplest discrete non-Abelian models and are the subject of many quantum-simulation proposals~\cite{Zohar_PRD2015,Bender_NJP2018, Fromm2023, Ballini_PRD2024, Ballini_Quantum2024,Gaz_PRR2025}.
They are also at the forefront of experimental efforts to implement states with non-Abelian topological order and topological quantum computing protocols~\cite{Iqbal2024,Lo_2025,Lo_2026,Sajith_PRXQ2026}.
Yet, their general phenomenology outside the weak coupling limit is largely unexplored, raising the questions of how deeply non-perturbative physics is imprinted into these models and whether they may provide a minimal setup to investigate the mechanisms underlying phenomena in QCD that are not yet fully understood, such as confinement.

\begin{figure*}
    \centering
     \includegraphics[width=\linewidth]{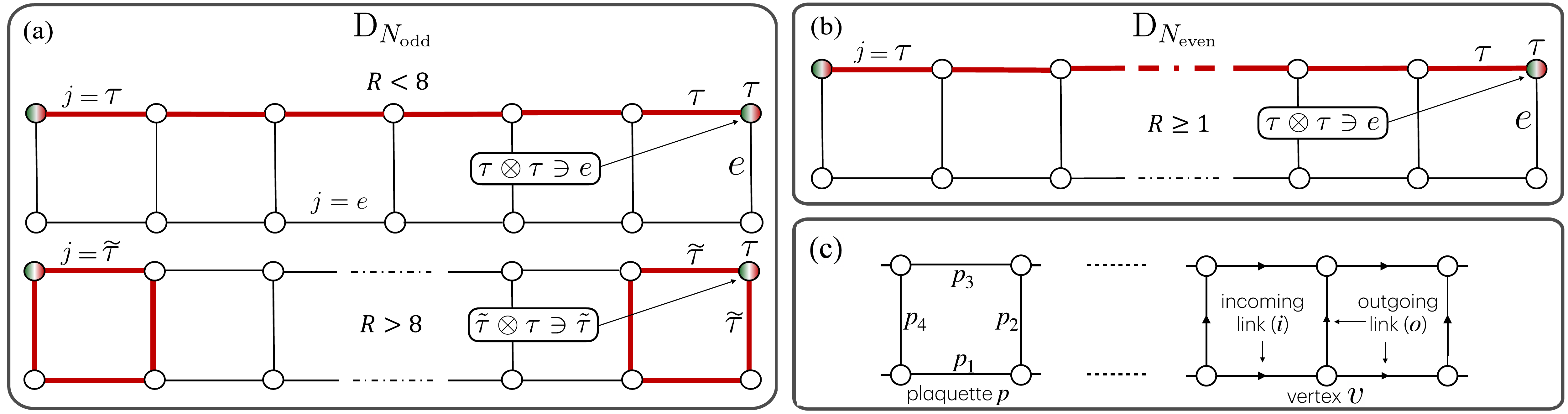}
    \caption{{\bf (a)} Ground-state configurations at strong coupling of $D_{N_{\rm odd}}$ LGTs on a ladder with static color charges on the corners. The string in the fundamental representation $\tau$ breaks when the charge separation $R$, in lattice units, is greater than 8, making the anti-fundamental $\taf-$glueballs energetically favorable. This threshold corresponds to the string and the gluelump configurations having the same number of links excited in the $\tau$ or $\tilde{\tau}$ irrep. The charge-glueball configuration is gauge-invariant due to the screening fusion rule of Eq.~\eqref{eq:fusion}.
    {\bf (b)} In $D_{N_{\rm even}}$, the GS at strong coupling always comprises the shortest $\tau$-string connecting the charged corners, since there is no equivalent to the screening fusion rule.
    {\bf (c)} Ordering of links in a plaquette operator and highlighting of incoming and outgoing links associated with a vertex of the directed lattice.}
    \label{fig:GSsketch}
\end{figure*}

We show that \new{$D_N$ groups allow for a precise analytical and numerical characterization of confinement at}  strong coupling \new{and its intimate connection} to the presence of a $\mathbb{Z}_2$ center.
For $N$ even, the group center is $\mathbb{Z}_2$ and a non-Abelian electric flux string connects---and confines---static probe charges over arbitrary distances (a well-known consequence of the nontrivial center subgroup~\cite{Greensite2003}). 
Conversely, when $N$ is odd, the center is trivial and \new{we reveal that} the string breaks {\em without matter production}, because gluon clouds screen the charges, forming bound states known in QCD as {\em gluelumps}~\cite{Greensite2003}, see Fig.~\ref{fig:GSsketch}(a-b) for a graphical representation. This string-breaking mechanism of fundamental charges for odd $N$, a consequence of the trivial center, does not have an analogue in SU(N) gauge theories.

A ladder geometry~\cite{Munk_PRB2018,Nyhegn2021,albert2021,Pradhan_PRB2024} allows for observing this drastically different behavior between even and odd $N$, combining both magnetic and electric field terms in a tractable 1D setting. In the following, we present tensor-network simulations of pure $D_N$ LGTs, analyzing the different strong-coupling behaviors of $D_3$ and $D_4$ symmetry groups.
A scaling analysis of the size of the $D_3$ gluelumps 
allows for a precise characterization of the continuum limit, where they retain a finite interaction strenght.
Our work demonstrates how discrete non-Abelian groups represent an ideal testbed to investigate the mechanisms behind non-perturbative phenomena such as confinement and their interplay with the mathematical structure of the theory, in a setting more accessible to classical and quantum simulations than QCD.

\emph{Model and fusion rules.---}We consider a ${D}_N$ pure-gauge theory in a ladder geometry, where each link hosts a Hilbert space $\mathcal{H}_{\rm link} = \{\ket{h},\:h\in D_N\}$, with lattice spacing $a$. To test confinement, we add two static color charges on the edges of the upper leg. Figure~\ref{fig:GSsketch} sketches the model and the relevant gauge-invariant configurations. The Kogut--Susskind Hamiltonian $H = H_B + H_E$ consists of two parts~\cite{kogut1975hamiltonian,Zohar_PRD2015}. The magnetic contribution 
\begin{equation}
  H_B = - \frac{c \hbar}{a g^2} \sum_p \mathfrak{R}[\textrm{Tr}(U^j_{p_1}U^j_{p_2}U^{j\dagger}_{p_3}U^{j\dagger}_{p_4})]  
\label{eq:plaquette_ham}
\end{equation}
 is the usual sum of plaquette operators, where ${g^2}$ is the dimensionless coupling constant and $U^j_{p_i}$ is the group connection in the $j-$th irreducible representation (irrep) acting on the $i-$th link of the $p-$th plaquette (see Fig.~\ref{fig:GSsketch}c). Hereafter, we consider the connection in the two-dimensional fundamental irrep $j=\tau$.  

The electric Hamiltonian
\begin{equation}\label{eq:h_el}
    H_E= \frac{c \hbar}{a} g^2 \sum_{l \in {\rm links}} \sum_j \alpha_j \Pi^j_l 
\end{equation} 
is the direct sum of one-body terms diagonal in the irreps, that is, the electric basis~\cite{Zohar_PRD2015,Mariani_PRD2023}.
 We consider the rescaling (canonical) transformation $H \to H' = \frac{a}{c \hbar} H$ to work in dimensionless, natural units for the energy, which, however, may still carry a scaling dimension with the lattice spacing $a$. 
 In these units, $g^2 \alpha_j$ is the electric energy of the $j-$th irrep and $\Pi^j_l$ the corresponding projector on the $l-$th link.

Physical quantities are invariant under local gauge transformations $\Theta_{h,v}$, $g\in D_N$.
In pure gauge theories, these read
\begin{equation}\label{eq:gauss_laws}
    \Theta_{h,v} = \prod_o \theta_{h,o}^L\prod_i\theta_{h,i}^R \ ,
\end{equation}
where $o$ ($i$) indicates the outgoing (ingoing) links connected to the vertex $v$ (see Fig.~\ref{fig:GSsketch}c), and $\theta^{L(R)}_h$ are the left (right) gauge transformations associated with the group element $h$. They act in the group-element basis as $\theta^L_h\ket{k}=\ket{hk}$ and $\theta^R_h\ket{k}=\ket{kh^{-1}}$. Gauge-invariant states in the neutral sector satisfy $\Theta_{h,v}\ket{\psi_{\textrm{phys}}} = \ket{\psi_{\textrm{phys}}}, \forall h, v$. 
These constraints determine how the irreducible representations surrounding the vertex can fuse, the so-called ``\textit{fusion rules}"~\cite{Simon_2023}, see Appendix~\ref{app:gauge_inv} for details.

The mathematical backbone of our results is the observation that fusion rules in $D_N$ groups work differently for even and for odd $N$. Finite groups allow for computing the Clebsch--Gordan (CG) coefficients and therefore, the fusion rules of representations, through their characters (see Appendix~\ref{app:fusion_rules_Dn}). Here, of central interest for us is that the following fusion rule, which we dub the screening fusion rule, holds only for odd $N$:
\begin{align}\label{eq:fusion}
\exists \ {\rm irrep} \ j \ |\quad j\otimes \tau \ni j\ ,
\end{align}
where $j$ is a non-Abelian irrep ($\dim(j)>1$). This rule means that the gauge field irrep $j$ is not changed as it flows through a fundamental charge $\tau$. In $D_N$ groups, Eq.~\eqref{eq:fusion} can be satisfied by the faithful anti-fundamental irrep $j=\taf$ (see Appendix~\ref{app:fusion_rules_Dn} for the definition) and the associated CG coefficients are determined by the group center~\footnote{The center $Z$ of a group $G$ is an Abelian subgroup that commutes with every other element of $G$, i.e., $Z = \{h \in G: \ \forall \ g \in G, \ hg=gh \}$. $Z$ is also a normal subgroup, meaning that $ghg^{-1} \in Z \ \forall \ g\in G, \ h \in Z$.}. For odd $N$, the center is trivial---the identity element $\{\mathbb{1}\}$---and the CG coefficient is one. For even $N$, the center is $\mathbb{Z}_2$ and the CG coefficient is zero.

The screening fusion rule has profound consequences for the physics of dihedral LGTs. Consider color charges that transform as the fundamental irrep of the group $\tau$. Then, in $D_{N_{\rm odd}}$ groups, the fusion rule of Eq.~\eqref{eq:fusion} allows them to be simultaneously the source and the sink of electric flux lines in the anti-fundamental representation $\taf$. At strong coupling and large distances, we thus expect static charges to be screened by the glueballs pinned at the charge positions. This is not the case for $D_{N_{\rm even}}$ groups, where Eq.~\eqref{eq:fusion} does not hold and a pair of static charges {\em must} be connected by a $\tau$-flux string. These different gauge-invariant configurations are sketched in Fig.~\ref{fig:GSsketch}(a-b).

\emph{DMRG results.---}To study the properties of the gluelumps at different couplings, we analyze the ground state of $D_3$ and $D_4$ LGTs with DMRG \new{(density-matrix renormalization group)}. Both groups have a single two-dimensional (fundamental) irrep, so that $\taf=\tau$. We consider the cases in which there are no charges $(q=0)$ and where the upper corners of the ladder are occupied by single static charges ($q=1)$.
All simulations are performed using the ITensor~\cite{ITensor,ITensor-r0.3} Julia library, with maximum MPS bond dimension $\eta=200$ and cutoff $\epsilon=10^{-11}.$  

\begin{figure}[t]
    \centering
    \includegraphics[width=\linewidth]{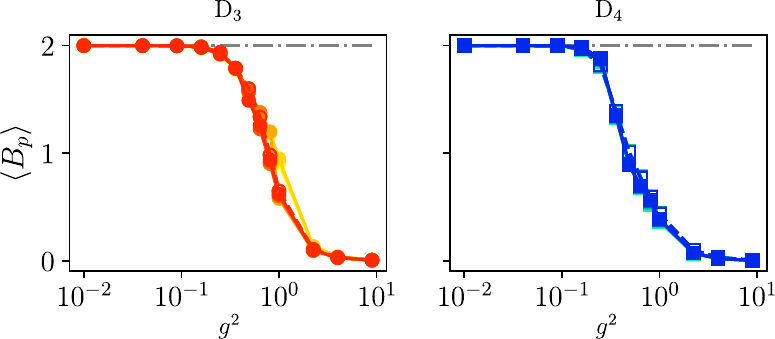}
    \caption{Magnetic energy density for ${\rm D}_3$ and ${\rm D}_4$ LGTs on a ladder. Continuous lines with filled markers indicate the presence of two static charges on the upper corners of the ladder, while dashed lines with empty markers stand for the neutral sector. 
    The results are almost independent of the number of rungs ($N_r \in [4,60]$, increasing for darker colors).
    The plaquette expectation value $\langle B_p \rangle \sim 2$ at $g^2\to 0$, corresponds to the absence of magnetic fluxes.}
    \label{fig:phase_diagram}
\end{figure}
In full (2+1)D geometries, discrete LGTs undergo a phase transition between a confined phase at strong coupling and a deconfined topologically-ordered phase at weak coupling~\cite{Fradkin_PRD1979,kitaev2002topological, Tantivasadakarn_PRL2023, Iqbal2024, Lo_2025}. On a ladder, however, the edges induce an effective electric field that prevents a phase transition at finite $g$~\cite{Nyhegn2021,Pradhan_PRB2024}.
Instead, the coupling drives a crossover between two smoothly-connected regimes that minimize either the electric or the magnetic energy.
This crossover is revealed by looking at the expectation value of the plaquette operators $B_p =\mathfrak{R}[\textrm{Tr}(U^j_{p_1}U^j_{p_2}U^{j\dagger}_{p_3}U^{j\dagger}_{p_4})]$, i.e., (minus) the magnetic energy density, as a function of $g^2$.
As seen in Fig.~\ref{fig:phase_diagram}, the phenomenology of $\langle B_p \rangle$ in $D_3$ and $D_4$ LGTs appears to be very similar. It vanishes at strong coupling, indicating a phase dominated by electric order. As $g^2\to 0$, it converges towards the character of the identity in the fundamental representation, $\chi^\tau(\mathbb{1})=2$. Physically, this value is associated with the absence of magnetic vortices and with the deconfined phase, which on a ladder is reduced to the point $g^2=0$.
The presence of the static charges mildly affects the magnetic excitations only in the crossover region.

\begin{figure}[t!]
    \centering
    \includegraphics[width=\linewidth]{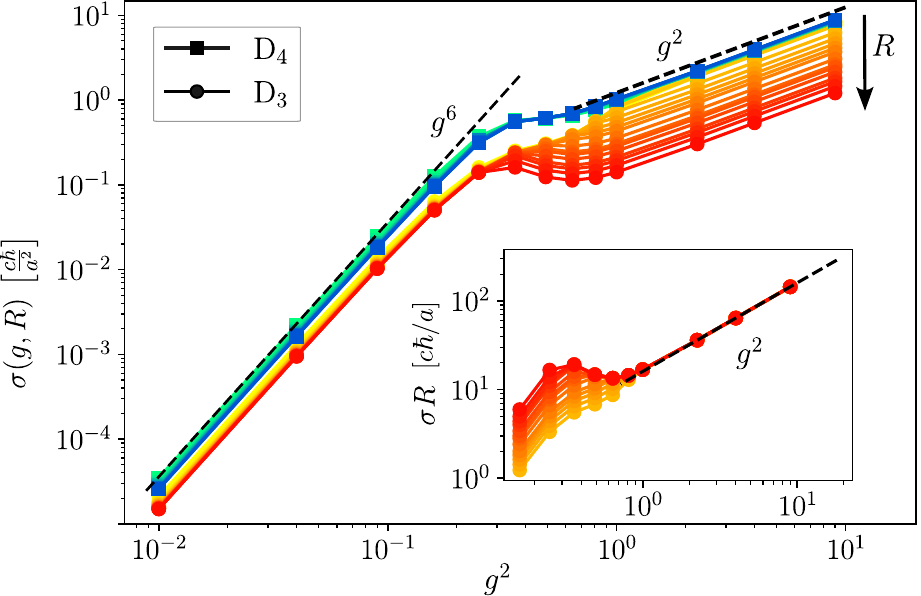}
     \caption{String tension as a function of the coupling $g^2$. 
     We compare different charge separations $R$, indicated by line shading, for $D_3$ (warm colors and circles) and $D_4$ (cold colors and squares). The weak confinement $\sigma \propto g^6$ at $g^2\lesssim 0.2$ is qualitatively similar for both models. 
     For $D_4$, the string tension $\sigma$ is independent of $R$, indicating the presence of a field string connecting the charges. 
     Instead, for $D_3$ at strong coupling, the string breaks, indicated by $\sigma$ decreasing with growing $R$.  
    The inset shows the energy difference $E_{q=1}(g,R) - E_{q=0}(g,R)=\sigma(g,R) R$ [see Eq.~\eqref{eq: string tension}]. Its independence of $R$ at strong coupling highlights the mutual screening of the two gluelumps, and $\sigma R=2M_0(g)$.}
    \label{fig:DeltaE_vsg}
\end{figure}

The effect of glueballs screening the static charges at strong coupling emerges clearly when comparing the string tension 
\begin{equation}\label{eq: string tension}
    \sigma(g,R) = \frac{E_{q=1}(g,R) - E_{q=0}(g,R)}{R}
\end{equation} 
for the two groups. $E_{q=1(0)}$ is the ground-state energy with (without) probe charges, and $R$ is their distance.
At strong coupling, the dominant contribution to the energy difference is linear with $R$, and $\sigma$ is proportional to $g^2(\alpha^\tau-\alpha^e)$. This is, indeed, what happens for $D_4$, where $\sigma$ shows no significant size scaling (see Fig.~\ref{fig:DeltaE_vsg}).
Conversely, at strong coupling $D_3$, the string tension decreases linearly as $R$ increases, making the energy difference $E_{q=1} - E_{q=0} = \sigma R$ independent of the charge separation, see inset of Fig.~\ref{fig:DeltaE_vsg}.
This asymptotic value $[E_{q=1}(g,R\to \infty) - E_{q=0}(g,R\to \infty)]= 2 M(g)$ can be interpreted as the renormalized contribution to the mass of the two gluelumps \footnote{Thanks to the fast convergence of the gluelump mass, we compute it in practice using a fixed large system size accessible to DMRG (here, $R=59$).}. Here, the fundamental (trivial) irrep has an energy $\alpha^\tau=1$ ($\alpha^e=0$), leading to $ M(g) \simeq M_0(g) = 4 g^2$, for $g^2 > 1$.

At weak coupling, both groups display \new{a trivial weak confinement due to the ladder geometry inducing an effective }background electric field, analogously to what can be observed in a $\mathbb{Z}_N$ theory~\cite{Nyhegn2021} (see Appendix~\ref{app:z2}). In such a model, second-order perturbation theory predicts the string tension to scale as $\sigma(g,R)\propto g^6$, which matches the behavior of the $D_N$ theories.
Table~\ref{tab: scaling} summarizes the behaviors of $D_3$ and $D_4$ in the two coupling regimes.
\begin{table}[]
    \centering
    \begin{tabular}{|c|c|c|}
    \hline
               & Strong coupling & Weak coupling  \\ \hline
         $D_3$ & $\min(8,R)g^2$          & $Rg^6$ \\ \hline
         $D_4$ & $Rg^2$          & $Rg^6$ \\ \hline
    \end{tabular}
    \caption{Summary of the scaling of the string energy $R\sigma (g,R)$ consistently with Eq.~\eqref{eq: string tension} for the two groups and the two regimes considered. $R$ is measured in lattice units $a$. }
    \label{tab: scaling}
\end{table}

\begin{figure}
    \centering
    \includegraphics[width=\linewidth]{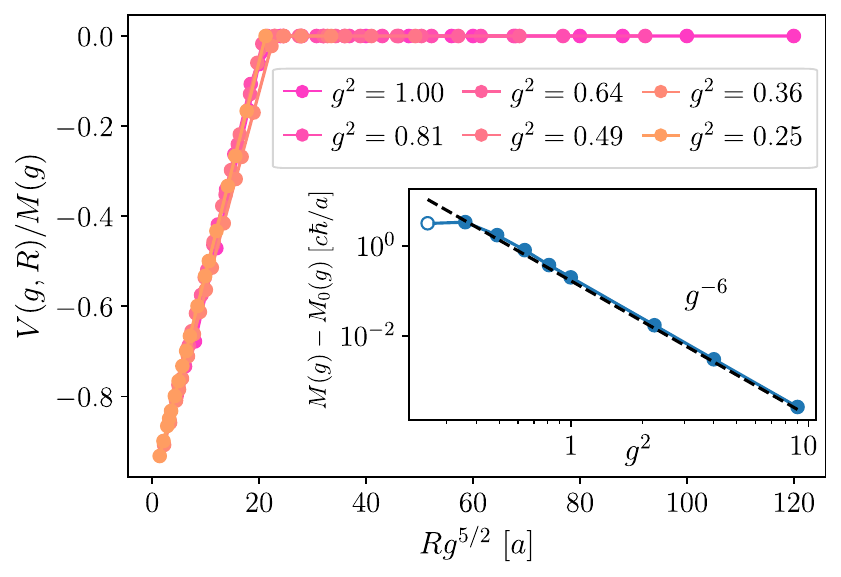}
    \caption{Rescaled interaction potential $V(g,Rg^\se)$, with $\se=5/2$ as a function of the separation between the two charges. The excellent collapse suggests that the screening distance scales as $R_s \sim g^{-5/2}$.
    Inset: deviation of the renormalized mass $M(g)$ from the behavior $M_0(g)=4g^2$ expected at strong-coupling. The leftmost point corresponds to a coupling strength where $R$ is too small to see the screening effect.}
    \label{fig:string_tension}
\end{figure}

The $R$-dependent crossover between the $g^2$ and the $g^6$ scaling in $D_3$ can be understood in terms of a growing effective size of the dressed particles as $g^2$ decreases, due to increasing electric-field fluctuations. 
The spread of the gluelumps along the ladder has a direct effect on their interaction range.
To characterize it, we study the scaling of the distance-dependent potential between two gluelumps, 
\begin{equation}
    V(g,R) = E_{q=1}(g,R) - E_{q=0}(g,R) -2 M(g) \ .
\end{equation}
When rescaling the distance $R$, see Fig.~\ref{fig:string_tension}, DMRG data obtained at different $g^2$ collapse onto a universal behavior of the rescaled potential $V(g,R g^{\se})/M(g)$, with $\kappa = 2.5(1) \simeq \frac{5}{2}$, showing a finite screening length $R_s$ at any nonzero coupling strength.
The mass follows $M(g)\simeq \gamma g^{-\me} - M_0(g)$, see inset of Fig.~\ref{fig:string_tension}, with $\me=5.99(5)\simeq 6$.

 These observations convey precious information regarding the continuum limit scaling, which is uniquely defined under a couple of assumptions. First, we request that the physical range $r_{\text{phys}}$ of the potential well is a finite length in the limit $a \to 0$.
 This assumption requires the lattice coupling $g$ to obey the scaling
 \begin{equation}
 \label{eq:gscaling}
  g \simeq \left( \zeta \frac{a}{r_{\text{phys}}} \right)^{1/\se}\,,
 \quad \mbox{for} \quad a \ll r_{\text{phys}}\,,
 \end{equation} 
 where the dimensionless value $\zeta = 21.2(4)$ is the $x$-axis location of the knee in Fig.~\ref{fig:string_tension}.
 According to this argument, since $g \propto a^{1/\se}$ and $\se>0$, the continuum limit is located at small $g$ values (as intuitively understood from the lattice extension of the gluelumps increasing with the strength of the plaquette term). 
 
 Second, we set the physical quasiparticle excitation energy $\mathcal{E}_{\text{phys}}$, related to one gluelump sitting at the edge, to be finite in the continuum limit. This fixes the scaling dimension of lattice unit energies, as  $M \simeq \gamma g^{-\me} \propto a^{-\me/\se}$\footnote{Here, we neglect the $M_0\propto g^2$ contribution assuming it is much smaller than the $\sim g^{-6}$ scaling at weak coupling.}, where we fitted
 $\gamma = 0.19(1) $ in Fig.~\ref{fig:string_tension}(inset). At the same time, $M(g)$ must scale linearly with $\mathcal{E}_{\text{phys}}$ (at fixed $r_{\text{phys}}$ and $a$) and be dimensionless, meaning we can write it as
  \begin{equation}
  M(g(a)) \simeq \xi \frac{\mathcal{E}_{\text{phys}} r_{\text{phys}}}{c \hbar} \left( \frac{r_{\text{phys}}}{a} \right)^{\me/\kappa}.
 \end{equation} 
 All that is left is to evaluate the dimensionless factor $\xi$, which, using $M \simeq \gamma g^{-\me}$ and Eq.~\eqref{eq:gscaling}, reads
  \begin{equation}
 \xi = \left(\frac{c \, \hbar}{ \mathcal{E}_{\text{phys}} \, r_{\text{phys}} \,
 }\right)
 \frac{\gamma}{\zeta^{\me/\se}}.
 \end{equation} 
 Thus, these two assumptions allowed us to perform the energy conversion from lattice to physical units in the continuum limit, similar to a running coupling analysis~\cite{Clemente_PRD2022,Crippa_CommPhys2025}.

Interestingly, the amplitude of the interaction potential $V$ between the two quasiparticles exhibits \emph{the same emergent scaling} as $M(g)$ with $g$, and therefore with $a$. This implies that in the continuum limit, this interaction potential will survive and be finite, thus providing measurable elastic scattering amplitudes for the field theory.

\emph{Comparison with SU($N$) LGT.---}
In SU($N$), the nontrivial center prevents the screening fusion rule for the fundamental representation, but not for other representations that have zero $N$-ality~\footnote{$N$-ality is related to the representation of the center subgroup $\mathbb{Z}_N$ in a given representation of SU($N$). Zero $N$-ality means that the center is represented by the identity matrix.}. 
One such example is the \new{adjoint irrep of SU(3) (${\bf 8}$), which fulfills the screening condition $\bf{8}\otimes \bf{8} \ni \bf{8}$, allowing for color-neutral objects composed by a ${\bf 8}$} charge screened by a gluon cloud, the QCD gluelumps~\cite{Greensite2003,Bali_PRD2004,Kratochvila_NucPhysB2003}. 
\new{Although gluelumps are not asymptotic states of QCD, they have been extensively studied via static sources, such as heavy quark--antiquark pairs, and are relevant for describing heavy mesons and quarkonium hybrids~\cite{Meyer2015,Berwein_PRD2015}, and for QCD extensions with gluino fields~\cite{Foster_PRD1999}.  Extensive numerical works allowed for the determination of their masses~\cite{Bali_PRD2004, Marsh_PRD2014, Herr_PRD2024,Ayala_JHEP2025}.}

In contrast to SU($N$), in dihedral groups particles in the fundamental representation can be part of the gluelumps. 
Hence, in a theory with dynamical matter, the charge components of the gluelumps are protected by a global U(1) symmetry. See Appendix~\ref{app:comparison} for details.

\emph{Conclusions.---} We have investigated the influence of fusion rules on the structure of gauge-invariant states in dihedral LGTs. \new{Only $D_{N_{\rm odd}}$ gauge symmetries allow for the screening of fundamental static charges at strong coupling due to gluelump formation,} i.e., bound states of a charge and a glueball. \new{A scaling analysis shows that the interaction strength} of such objects remains finite in the continuum limit, raising the perspective of studying related scattering phenomena.
\new{Moreover, it suggests the possibility of deriving an effective field theory with $D_N$ anyons, relevant for quantum wires and topological quantum computing~\cite{albert2021}.}

In $D_{N_{\rm even}}$, {\new gluelumps are forbidden} by a (nontrivial) $\mathbb{Z}_2$ center.  
The same property also influences the local-fermionic-parity conservation when the theory includes dynamical fermions: only when $\mathbb{Z}_2$ is a normal subgroup of the gauge group, the system maps onto a local Hamiltonian without fermionic degrees of freedom in arbitrary dimensions~\cite{Zohar_PRB2018, Zohar_PRD2019,Gaz_PRR2025,Zohar_PRD2015,Felser_PRX2020,Cataldi_PRR2024} (see also Appendix~\ref{app:parity}).

The group-theoretic origin of $D_{N_{\rm odd}}$ gluelumps guarantees that they remain valid gauge-invariant states in any spatial dimension, whereas their scaling behavior and the spatial structure of the gluon cloud are specific to the ladder geometry.
This raises interesting questions about the gluelump stability in the presence of a finite-coupling deconfinement phase transition in larger-dimensional lattices.
A complete picture also requires understanding the emergence of gluelump-mediated string breaking out of equilibrium. Of particular interest is its coexistence with dynamical charges, which provide a further (competing) screening mechanism.

Our results showcase how discrete non-Abelian LGTs already display much of the interesting phenomenology typically associated with Lie groups, such as SU(3) for QCD, making them an ideal testbed for quantum- and quantum-inspired simulation protocols.

{\em Note added.---}After the submission of this work, a trapped-ion quantum simulation of string breaking by gluelump formation has been reported~\cite{john2026nonabelianstring} for static charges in the adjoint irrep of SU(2)---the spin-$1$ representation.

\begin{acknowledgments}
{\em Acknowledgments.---}E.B. and M.B. thank Giuseppe Clemente for useful discussions.
This project has received funding from the European Union
via Horizon Europe research and innovation programme under grant agreement No 101080086 NeQST,
and via Cascade calls of the Extended Partnership NQSTI $-$ project OPTIMISTIQ.
P.H. has further received funding from the Swiss State Secretariat for Education, Research and Innovation (SERI) under contract number UeMO19-5.1, from the QuantERA II Programme through the European Union’s Horizon 2020 research and innovation programme under Grant Agreement No 101017733, from the European Union under NextGenerationEU, PRIN 2022 Prot. n. 2022ATM8FY (CUP: E53D23002240006), from Fondazione Cassa di Risparmio di Trento e Rovereto (CARITRO) through the project SQuaSH - CUP E63C24002750007.
Views and opinions expressed are however those of the author(s) only and do not necessarily reflect those of the European Union or the European Commission. Neither the European Union nor the granting authority can be held responsible for them.
This work was supported by Q@TN, the joint lab between the University of Trento, FBK—Fondazione Bruno Kessler, INFN—National Institute for Nuclear Physics, and CNR—National Research Council.

P.P.P. acknowledges support from: 
MCIN/AEI (PGC2018-0910.13039/501100011033,  CEX2019-000910-S/10.13039/501100011033, Plan National STAMEENA PID2022-139099NB, project funded MCIN and  by the “European Union NextGenerationEU/PRTR" (PRTR-C17.I1), FPI); Ministry for Digital Transformation and of Civil Service of the Spanish Government through the QUANTUM ENIA project call - Quantum Spain project, and by the European Union through the Recovery, Transformation and Resilience Plan - NextGenerationEU within the framework of the Digital Spain 2026 Agenda; CEX2024-001490-S [MICIU/AEI/10.13039/501100011033]; Fundació Cellex;
Fundació Mir-Puig; Generalitat de Catalunya (European Social Fund FEDER and CERCA program; Barcelona Supercomputing Center MareNostrum (FI-2023-3-0024); 
Funded by the European Union (HORIZON-CL4-2022-QUANTUM-02-SGA, PASQuanS2.1, 101113690, EU Horizon 2020 FET-OPEN OPTOlogic, Grant No 899794, QU-ATTO, 101168628),  EU Horizon Europe Program (No 101080086 NeQSTGrant Agreement 101080086 — NeQST).

\end{acknowledgments}

\onecolumngrid
\appendix

\section{Group structure of $D_N$ and conjugacy classes}
The dihedral group $D_N$ has order $2N$, and it represents the symmetry group of an equilateral polygon with $N$ sides. A set of generators for the group consists of the rotation of $2\pi/N$ ($r$) and a reflection ($s$): 
\begin{align}
    D_N = \{\mathbb{1},r,r^2,\dots,r^{N-1},s,sr,sr^2,\dots,sr^{N-1}\}\ .
\end{align}
Two elements of a group $a,b\in G$ are called conjugate if there exists a group element $g\in G$, such that $a = gbg^{-1}$. Since conjugacy defines an equivalence relation, the group factorizes into conjugacy classes, each of which contains mutually conjugate group elements. In the case of the dihedral group, the conjugacy classes depend on whether $N$ is even or odd. Specifically, 
\begin{align}
    D_{N_{\rm even}} = &\{\mathbb{1}\} \cup\{r,r^{-1}\}\cup\{r^2,r^{-2}\}\cup\dots\notag\\
    &\dots\cup\{r^{N/2-1},r^{-N/2+1}\}\cup\{r^{N/2}\}\\
    &\cup\{s,sr^2,sr^4,\dots,sr^{N-2}\}\cup\{sr,sr^3,\dots,sr^{N-1}\} \notag \ ,
\end{align}
and 
\begin{align}
D_{N_{\rm odd}} = &\{\mathbb{1}\} \cup\{r,r^{-1}\}\cup\{r^2,r^{-2}\}\cup\dots\notag \\
&\dots\cup\{r^{(N-1)/2},r^{-(N-1)/2}\} \\
&\cup\{s,sr,sr^2,sr^3,sr^4,\dots,sr^{N-2},sr^{N-1}\} \notag \ .
\end{align}
Here, we used the notation $r^{-k}$ for the rotation of $-2\pi k/N$. Crucially, the element $r^{N/2}$, which corresponds to a $\pi$-rotation, is a group element only for even $N$ and has its own conjugacy class. This is a direct consequence of $D_{N_{\rm even}}$ having a $\mathbb{Z}_2$ center, which is precisely $\{\mathbb{1},r^{N/2}\}$. Each element of the center builds its own conjugacy class by definition.

\section{ Irreducible representations and fusion rules of $D_N$}\label{app:fusion_rules_Dn}
The simple structure of $D_N$ groups allows for analytic calculations of fusion rules, in terms of the decomposition of an arbitrary representation into irreducible ones.
In finite groups, the number of conjugacy classes is equal to the number of irreducible representations. This means, $D_N$ has in total $N/2+3$ for even $N$ and $(N-1)/2+2$ for odd $N$ inequivalent irreducible representations. Furthermore, the cardinality of the group is connected to the dimensions of the irreducible representations
\begin{align}
    \sum_j \dim^2(j) = |D_N| = 2N\, .
\end{align}
For dihedral groups, each representation is either one- or two-dimensional. 

One can characterize (up to an isomorphism) the two-dimensional irreducible representations of $D_N$ as follows. Let $\omega = e^{2\pi i/N}$ and $h \in \{1,\dots,N/2-1\}$ for $N$ even and $h \in \{1,\dots,(N-1)/2\}$ for $N$ odd. 
The irreducible representation $\tau_h$ of the group elements is given as follows:
\begin{align}
    D^{\tau_h}(r^k) = \begin{pmatrix}
\omega^{hk} & 0\\
0 & \omega^{-hk} 
\end{pmatrix}, \:\: D^{\tau_h}(sr^k) = \begin{pmatrix}
0 & \omega^{-hk}\\
\omega^{hk} & 0
\end{pmatrix} \ .
\end{align}
Here, we have chosen the form of the irreps, for which $\overline{\tau_h}(g)$ and $\tau_h(g^{-1})$ are equivalent up to reordering. Therefore, we can drop the conjugation when dealing with fusion rules.
The characters are then
\begin{align}
    &\chi_{\tau_h}(sr^k) = \chi_{\tau}(sr^k) = 0\ ,\notag\\
    &\chi_{\tau_h}(\mathbb{1}) = \chi_{\tau}(\mathbb{1}) = 2\ ,\notag\\
    &\chi_{\tau_h}(r^k) = 2\cos\bigg(\frac{2\pi hk}{N}\bigg)\, .
\label{eq:characters_D_N}
\end{align}
In the following, we will denote the fundamental representation $\tau_{h=1}$ simply as $\tau$.

By the great orthogonality theorem for finite groups, each representation $J$ of $G$ can be written as a direct sum over irreducible representations 
\begin{align} \label{decomp}
    J = \bigoplus_{j \: \mathrm{irrep}} j^{\otimes a_j}\, ,
\end{align}
where $a_j$ is the multiplicity of the $j$-th irreducible representation. The multiplicity can be calculated by referring to the characters of the group in each irrep. 
When $a_j=0$, the corresponding irrep does not appear in the direct sum composing $J$. 
The multiplicities can be computed as 
\begin{align}
    a_j = (\chi_j,\chi_J) &:= \frac{1}{|G|}\sum_{g\in G}\overline{\chi_j(g)}\chi_J(g)\notag\\&\equiv \frac{1}{|G|}\sum_{C }|C|\overline{\chi_j(C)}\chi_J(C) \ , 
\end{align}
where $\chi_J(g)$ is the character of element $g$ in $J$ and $\chi_j(g)$ is its character in $j$. The bar means complex conjugation. The second equality holds due to the fact that characters are invariant over conjugacy classes, labeled by $C$.

We now want to demonstrate that the fusion rule in Eq.~\eqref{eq:fusion} holds for odd $N$ but not for even $N$. 
Since the $D_N$ characters are real, thus invariant under conjugation, it can be rewritten as $\tau_{h} \otimes \tau_{h} \ni \tau$. 
That is, we aim to show that the decomposition of $J=\tau_{h} \otimes \tau_{h}$ as per Eq.~\eqref{decomp} into a direct sum yields the multiplicity associated to the fundamental irrep  as  $a_{\tau}=0$ for $N$ even, while $a_{\tau}=1$ is possible for $N$ odd. 

To prove the above statement, we consider the more general case $J = \tau_h^{\otimes l}$, where $l$ is even. To find the multiplicity $a_{\tau}$, we need to compute the inner product 
\begin{align}
(\chi_{\tau},\chi_{\tau_h^{\otimes l}}) = \frac{1}{|D_N|}\sum_{C}|C|\overline{\chi_{\tau_h}(C)}^l\chi_{\tau}(C) \ ,
\end{align}
where we used the fact that $\chi_{\tau_h^{\otimes l}} = \chi_{\tau_h}^l$.

Let us begin with $N$ even. In this case, the inner product becomes 
\begin{align}
|D_N|(\chi_{\tau},\chi_{\tau_h^{\otimes l}}) =& \chi_{\tau}(\mathbb{1})\chi_{\tau_h}^l(\mathbb{1}) + \chi_{\tau}(r^{N/2})\chi_{\tau_h}^l(r^{N/2}) \notag\\&+2\sum_{k = 1}^{N/2-1}\chi_{\tau}(r^k)\chi_{\tau_h}^l(r^k)\notag\\&+(N/2)\chi_{\tau}(s)\chi_{\tau_h}^l(s)\notag\\&+ (N/2)\chi_{\tau}(sr)\chi_{\tau_h}^l(sr) \ .
\label{eq:inner_product_D_even}
\end{align}

Substituting the value of the characters of Eq.~\eqref{eq:characters_D_N} into Eq.~\eqref{eq:inner_product_D_even}, one gets
\begin{align}
|D_N|(\chi_{\tau},\chi_{\tau_h^{\otimes l}}) = &2^{l+1} + 2^{l+1}(-1)^{hl+1}\notag\\+&2^{l+2}\sum_{k=1}^{N/2-1}\cos^l\bigg(\frac{2\pi hk}{N}\bigg)\cos\bigg(\frac{2\pi k}{N}\bigg) \ .
\end{align}
The second term has an overall negative sign, since $l$ is even, and cancels exactly the first term. These two are associated with the conjugacy classes of the $\mathbb{Z}_2$ center $Z=\{ \mathbb{1}, r^{N/2} \}$.
The third term can be shown to be identically zero for all allowed $h$. Here, we need to look at two different cases separately. If $N/2-1$ is even, then the number of summands in the sum is even, and they cancel pairwise. For $N/2-1$ odd, the number of summands in the sum is odd, and all but one summand cancel pairwise. The remaining summand, corresponding to $k=N/4$, is, however, zero, since $\cos(\pi/2) = 0$ and therefore the sum vanishes also in this case. Overall, for $N$ even, the coefficient $a_{\tau} = 0$.

When $N$ is odd, the $\pi$ rotation is not a group element, and the contribution from the identity conjugacy class will not be canceled. 
The inner product becomes 
\begin{align}
|D_N|(\chi_{\tau},\chi_{\tau_h^{\otimes l}}) =& \chi_{\tau}(\mathbb{1})\chi_{\tau_h}^l(\mathbb{1})\notag\\&+2\sum_{k = 1}^{(N-1)/2}\chi_{\tau}(r^k)\chi_{\tau_h}^l(r^k)\notag\\&+N\chi_{\tau}(s)\chi_{\tau_h}^l(s) \ .
\label{eq:inner_product_D_odd}
\end{align}
Substituting the character values yields
\begin{align}\label{eq:char_odd}
|D_N|(\chi_{\tau},\chi_{\tau_h^{\otimes l}}) &= 2^{l+1}\notag\\ &+ 2^{l+2}\sum_{k=1}^{(N-1)/2}\cos^l\bigg(\frac{2\pi hk}{N}\bigg)\cos\bigg(\frac{2\pi k}{N}\bigg)\ .
\end{align}
We need to evaluate the second term in the last equation and show that there are $h$ and $l$, for which the inner product is not zero. For example, in the case of $N = 3$, which we treat in the main text, for $l = 2$, and $h = 1$ (i.e., the only two-dimensional, irrep $\tau_{h=1}\equiv\tau$) it holds 
\begin{align}
    |D_3|(\chi_{\tau},\chi_{\tau_1^{\otimes l}}) = 2^{2+1}+2^{2+2}\bigg(-\frac{1}{8}\bigg) = 6
\end{align}
and therefore $a_{\tau} = (\chi_{\tau},\chi_{\tau^{\otimes 2}}) = 1$. 

For odd $N>3$, the fundamental representation $\tau$ does not satisfy Eq.~\eqref{eq:fusion} any longer ---$(\chi_\tau, \chi_{\tau^{\otimes 2}})=0$--- while the faithful representation with the largest $h = (N-1)/2$ does. This is the anti-fundamental representation, denoted by $\taf$. Indeed, setting $h=(N-1)/2$ and $l=2$ in Eq.~\eqref{eq:char_odd} leads to
\begin{align}
(\chi_{\tau},\chi_{\taf^{\otimes 2}}) = 1
\end{align}
for all odd $N$. This means that the state on a charged vertex, where two semilinks are in the $\taf$ representation, is gauge invariant, satisfying the screening fusion rule of Eq.~\eqref{eq:fusion} in the main text.

\begin{figure}
    \centering
    \includegraphics[trim={5cm 20cm 5cm 4cm 0},clip,width=0.5\linewidth]{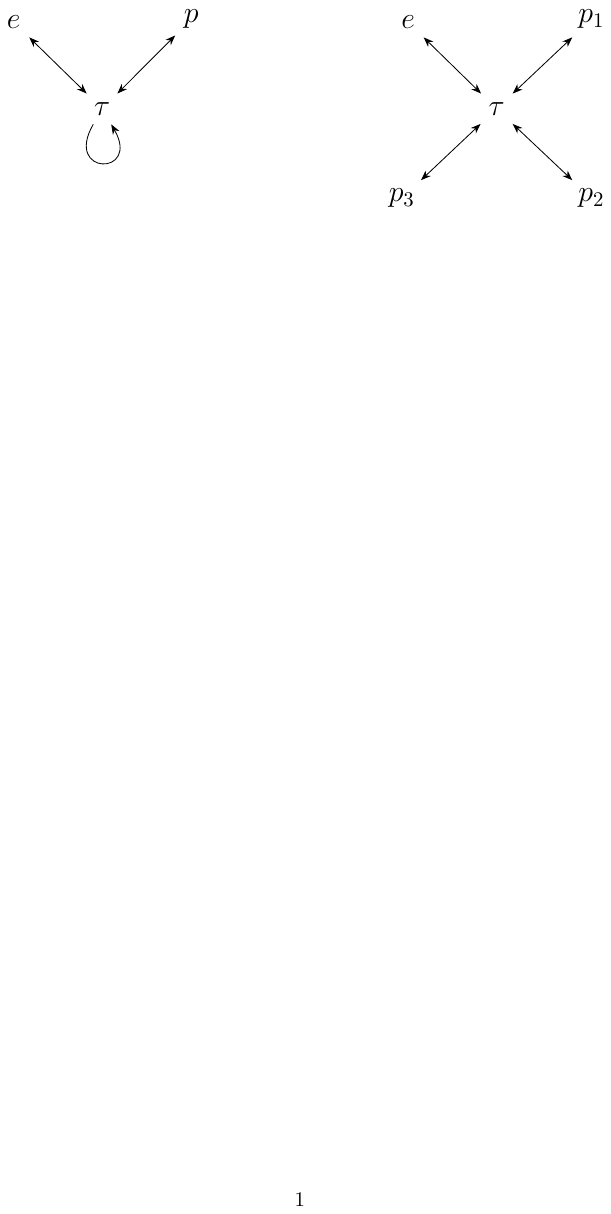}
    \caption{Graphic comparison of the fusion rules of $D_3$ (left) and $D_4$ (right). Arrows indicate nonvanishing Clebsch--Gordan coefficients. For $D_3$, the fundamental representation $\tau$ has a non-vanishing Clebsch--Gordan coefficient with itself, while for $D_4$, only terms connecting $\tau$ with the one-dimensional representations are finite.}
    \label{fig: fusion rules}
\end{figure}

In summary, for $N$ even, we have $\tau_h\otimes\tau_h \not\ni \tau$. In contrast, for $N$ odd, $\tau_h\otimes\tau_h = \dots\oplus  \tau$ when $h=(N-1)/2$. Hence, when two copies of a non-Abelian representation for $N$ odd mix, the result can be in the fundamental representation, while this is not the case for $N$ even.
The fusion rules of $D_3$ and $D_4$, the groups considered in the main text, are summarized in Fig.~\ref{fig: fusion rules}.

Similar calculations suggest that $\taf$-gluelumps are dynamically connected to the $\tau$-flux string connecting the two static charges. To show that, one has to evaluate the following inner product
\begin{align}
&|D_N|(\chi_{\tau}^{\otimes l+1},\chi_{\taf}) = \\ 
&\quad=2^{l+2}+ 2^{l+3}\sum_{k=1}^{(N-1)/2}\cos\bigg(\frac{\pi k(N-1)}{N}\bigg)\cos^{l+1}\bigg(\frac{2\pi k}{N}\bigg) \ . \notag
\end{align}
The smallest integer $l$, for which the inner product $(\chi_{\tau},\chi_{\taf^{\otimes 2}})$ is one, is 
\begin{align}
    l = \begin{cases}
    0, \:\: N = 3\\
    \frac{N-3}{2}, \:\: N>3 \ .
    \end{cases}
\end{align}
Physically, it implies that we need to apply the group connection $U^\tau$, as part of plaquette operators, $l(N)$ times, to transit from the fundamental string to the $\taf$ loops---the dynamical string breaking becomes a process of higher order as $N$ increases, making it non-perturbative in the limit of odd $N\rightarrow\infty$.

\section{Gauge transformations, static charges, and fusion rules}\label{app:gauge_inv}
The decomposition $J = \bigoplus_{j \: \mathrm{irrep}} j^{\otimes a_j}$ of general representations into irreducible representations has direct implications in the description of the ladder in the presence of static charges at its corners.
Let us consider the case of the top left corner in Fig.~1(a) of the main text (the static charge at the opposite corner has an analogous behavior). In the presence of a static charge $\tau$, there is a multiplet of physical states $\ket{\Psi_{{\rm phys},m}}$ ($m=1,\dots,\dim(\tau)$) transforming as the $\tau$ irreducible representation under the local gauge transformation $\Theta_{h,v_\tau}$, where $v_\tau$ labels the top left corner,
\begin{equation} \label{staticchargetr1}
\theta_{h,r}^R\theta^L_{h,l}\ket{\Psi_{{\rm phys},m}} = \sum_n D^\tau_{n,m}(h)\ket{\Psi_{{\rm phys},n}}\,,
\end{equation}
where $r$ labels the first rung, $l$ is the first link in the upper leg, and $ D^\tau_{n,m}(h)$ is the $2\times 2$ matrix that represents the group element $h$ in the fundamental representation $\tau$.
To describe  $\ket{\Psi_{{\rm phys},m}}$, we consider explicitly its decomposition in terms of the states $\ket{j_i,m_i,n_i}$ of the links $i=r$ and $i=l$ and the state $\ket{\tau,m}$ of the static charge: 
\begin{equation}
\ket{\Psi_{{\rm phys},m}} = \sum_{j_r,j_l,n_r,m_l} A^{j_r,j_l}_{n_r,n_l} \ket{j_r,m_r,n_r}\ket{j_l,m_l,n_l}\ket{\tau,m}\,.
\end{equation}
The left hand side of Eq. \eqref{staticchargetr1} becomes equivalent to:
\begin{align}\label{staticchargetr2}
\theta_{h,r}^R\theta^L_{h,l} & \sum_{j_r,j_l,n_r,m_l} A^{j_r,j_l}_{n_r,n_l} \ket{j_r,m_r,n_r}\ket{j_l,m_l,n_l}\ket{\tau,m} = \nonumber \\
& \sum_{j_r,j_l,n_r,m_l} A^{j_r,j_l}_{n_r,m_l} D^{j_r}_{n_r',n_r}(h^{-1}) D^{j_l}_{m_l,m_l'}(h)
\ket{j_r,m_r,n_r'}\ket{j_l,m_l',n_l}\ket{\tau,m} \ .
\end{align}
Based on the Clebsch--Gordan series relation~\cite{Brink1993},
\begin{equation} \label{CBseries}
D^{j_r}_{n_r',n_r}(h^{-1}) D^{\tau}_{m',m}(h^{-1})
=\sum_{I,n,n'} \bracket{j_r n_r' \tau m'}{I n'} \bracket{I n}{j_r n_r \tau m} D^{I}_{n'n}(h^{-1})\ ,
\end{equation}
Eq.~\eqref{staticchargetr2} is equivalent to the right-hand side of Eq.~\eqref{staticchargetr1} only if the coefficients $A$ are of the form~\cite{Zohar_NJP2016}
\begin{equation}
A^{j_r,j_l}_{n_r,n_l} = \alpha(j_r,j_l) \left\langle j_r n_r \tau m | j_l m_l  \right \rangle\,,
\end{equation}
where $\alpha$ depends on the irreducible representations $j_l,j_r$ only and $\left\langle j_r n_r \tau m | j_l m_l  \right \rangle$ is the conjugate of a Clebsch--Gordan coefficient. The previous equation states that the ingoing representations $j_r$ and $\tau$ must fuse into $j_l$, such that
\begin{equation}
\tau \otimes j_r \ni j_l \ .
\end{equation}
To realize the gluelump state, it is necessary that both links are in the same irreducible representation and combine with the charge into a group singlet, which is verified for $j_r = j_l = \taf$ and results in Eq.(4) of the main text. 
An analogous construction holds when we consider a vertex with three links in the bulk of the ladder. In this case, $\tau$ is replaced by the irreducible representation characterizing the states of the additional link (see the construction for 2D lattices in Ref.~\cite{Zohar_NJP2016}).

\section{Rishons and gauge invariant basis}
\label{app:rishons}

\begin{figure}[t!]
    \centering
    \includegraphics[width=0.8\linewidth]{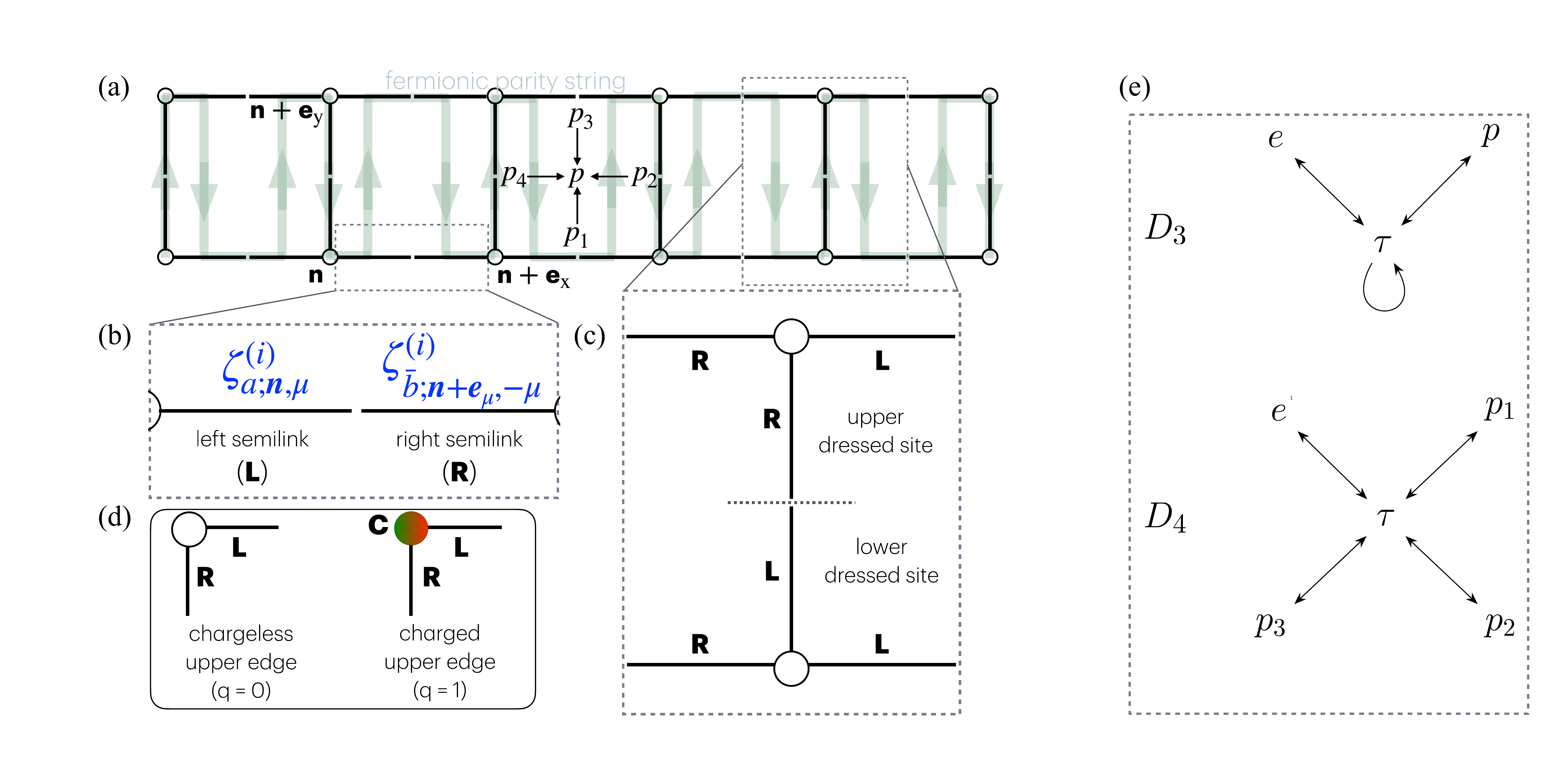}
     \caption{(a) Sketch of the ladder geometry, highlighting the plaquette operator $H_B = - \frac{c \hbar}{a g^2} \mathfrak{R}[\textrm{Tr}(U^j_{p_1}U^j_{p_2}U^{j\dagger}_{p_3}U^{j\dagger}_{p_4})]  $. The shaded path indicates the order in which we define the fermionic rishon parity in Sec.~\ref{app:parity}.
     (b) Splitting of a gauge field living on a link into two rishon operators.
     (c) Dressed sites of the ladder consisting of three semilinks. Sites on the upper legs have two ingoing R- and one outgoing L-semilinks, while sites on the lower legs have the opposite.
     (d) The upper corners of the ladder are where we insert static charges to probe confinement.
     (e) Graphic comparison of the fusion rules of $D_3$ (left) and $D_4$ (right). Arrows indicate nonvanishing Clebsch--Gordan coefficients. For $D_3$, the fundamental representation $\tau$ has a non-vanishing Clebsch--Gordan coefficient with itself, while for $D_4$, only terms connecting $\tau$ with the one-dimensional representations are finite.}
    \label{fig:semilinks}
\end{figure}

In this Appendix, we describe in detail the local gauge invariant basis used for the numerical calculation and how we used it to represent the gauge transformations and the Hamiltonian.

\subsection{Local dressed basis in rishon formalism}\label{app:basis}

We proceed with a reformulation of the $D_N$ lattice gauge theory in a local dressed sites basis, which allows for efficient Tensor Network implementation~\cite{Zohar_NJP2016,Felser_PRX2020,magnifico_natcom2021,Cataldi_PRR2024,Calajo_PRXQ2024}. We solve the non-Abelian Gauss's law on each vertex of the lattice by splitting each link into two halves, each attached to its adjacent vertex. This procedure can be reversed once the so-called selection rules are imposed on semilinks that correspond to the same link. 

In the electric basis, the parallel transporter $U_{ab}$, in the fundamental representation $\tau$, can be expressed as~\cite{Zohar_PRD2015}
\begin{equation}\label{eq:u_rep}
 \begin{aligned}
   \bra{{j_1; m_1, n_1}}U_{ab} \ket{j_2; m_2, n_2} 
   &\quad =  \sqrt{\frac{\textrm{dim}(j_2)}{\textrm{dim}(j_1)}}  \braket{j_2, m_2; \tau, a}{j_1,m_1}^{*}
   \braket{j_2, n_2; \tau, b}{j_1,n_1}
   \\
   &\quad = \sqrt{\textrm{dim}(j_1) \textrm{dim}(j_2)} 
   \left( \begin{array}{ccc}
    j_2 & \tau & \bar{j}_1 \\
    m_2 & a &  \bar{m}_1
   \end{array} \right)^{*}
   \left( \begin{array}{ccc}
    j_2 & \tau & \bar{j}_1 \\
    n_2 & b & \bar{n}_1
   \end{array} \right) \\
    &\quad 
     = {\braket{\tau, a;\bar{j}_1,\bar{m}_1}{\bar{j}_2, \bar{m}_2}}^{*}
   \braket{j_2, n_2; \tau, b}{j_1,n_1}  \ , 
 \end{aligned}
\end{equation}
where we explicitly express the fusion rules via the Clebsch--Gordan coefficients, or $3J$-symbols.
We can also explicitly carry out the calculations and get~\cite{Zohar_PRD2015}
\begin{align}
         \bra{{j_1; {m}_1, n_1}}U_{ab} \ket{j_2; {m}_2, n_2} =  
        \frac{\sqrt{\textrm{dim}(j_1)\textrm{dim}(j_2)}}{|G|} \sum_{g \in G} D^\tau_{ab}(g)D^{j_1*}_{m_1 n_1}(g)D^{j_2}_{m_2 n_2}(g) \ ,
\end{align}
\noindent with $D^j_{\alpha \beta}(g)$ denoting the matrix elements of the group element $g$ in the irreducible representation $j$. Equation~\eqref{eq:u_rep} shows that the group connection fuses separately the two indices of multi-dimensional representations; for instance, the $n_2$ and $m_2$ enter two different Clebsch--Gordan coefficients.
Therefore, each link can be split into left and right semilinks, such that $\ket{j,m_L,m_R}= \ket{j_1,m_L}\otimes\ket{j_2,m_R}\delta_{j_1,j_2}\delta_{j,j_1}$, where $\delta_{j_1,j_2}$ fix the same representation on the two semilink. 
Then, the parallel transporter can be split into rishon operators~\cite{Brower1999,Banerjee2013,Silvi2017}, distinctly acting on the left or the right semilink Hilbert space  [see Fig.~\ref{fig:semilinks}(b)],
\begin{equation}\label{eq:rishon_decomposition}
    U_{ab; \textbf{n}, \textbf{n}+\textbf{e}_{\mu}} = \sum_i \zeta_{a;\textbf{n},\mu}^{(i)} \zeta_{\bar{b}; \textbf{n}+\textbf{e}_{\mu},-\mu}^{(i)\dagger} \ ,
\end{equation}
where the index $i$ labels different rishon modes. 
We make explicit the action of the parallel transporter connecting the sites $\textbf{n}$ and $\textbf{n}+\textbf{e}_{\mu}$ ($\mu = \rm x,\rm y$), and the tensor product is projected onto the space where left and right rishons are in the same irrep. The label $\pm \mu$ indicates the position, with respect to the site $n$, of the semilink the rishon operator acts on.
Splitting $U_{ab}$ gives the possibility to define gauge transformations and the Hamiltonian in terms of dressed sites, i.e., a site $\textbf{n}$ with the attached semilinks.

In non-Abelian groups, left- and right-multiplication operators can be distinguished, in the group element basis, by the following definition:
\begin{align}
    &\theta^L_h \ket{g} = \ket{hg} \ , \nonumber \\
    &\theta^R_h\ket{g} = \ket{gh^{-1}} \ .
\end{align}
These operators appear in the gauge transformation $\Theta_{h,v} = \prod_o \theta_{h,o}^L\prod_i\theta_{h,i}^R$. Once split on the rishon spaces, the action of $\theta_g^{L(R)}$ operators is non-trivial only on the outgoing (ingoing) semilink:
\begin{equation}
    \theta^L_g = \Lambda^L_g \otimes \mathbb{1}_d \ , \quad \theta^R_g=\mathbb{1}_d\otimes \Lambda^R_g \ ,
\end{equation}
with $\mathbb{1}_d$ the identity acting on the $d$-dimensional rishon space.
On a ladder, gauge transformations on a vertex are given by the product of three $\Lambda$ transformations. This indicates that for a given dressed site, i.e., the vertex and the three semilinks, a gauge invariant basis can be obtained by removing states that transform non-trivially under the product of $\Lambda$ transformations. Due to the choice of orientation of the links, one has to differentiate between upper and lower vertices on the ladder, as well as between the different edges. As one can see in Fig.~\ref{fig:semilinks}(c), the lower dressed sites are constructed by taking the tensor product of one right (R) and two left (L) semilinks, and the upper dressed sites by two right (R) and one left (L) semilink. 

In the following, we write the operators and the transformations aforementioned for the $D_3$ case.
We label the representation basis for $D_3$ as $\{ \ket{0}, \ket{p}, \ket{rr}, \ket{rg}, \ket{gr}, \ket{gg} \}$, where $\ket{0}$ indicates the trivial representation, $\ket{p}$ the parity, and the states in the fundamental representation $\ket{\tau ab}$ are labeled by expressing internal indices (colors) as ``red" and ``green". Thus, the parallel transporter $U_{ab}$ reads
\begin{equation}
\label{eq:par_tranD3}
    U_{ab} = \left( 
    \begin{array}{c|c|cccc}
         0 & &  \frac{1}{\sqrt{2}}d_{rr} & \frac{1}{\sqrt{2}}d_{rg} & \frac{1}{\sqrt{2}}d_{gr} & \frac{1}{\sqrt{2}}d_{gg} \\ \hline
         & 0 & \frac{1}{\sqrt{2}}d_{rr} & \frac{-1}{\sqrt{2}}d_{rg} & \frac{-1}{\sqrt{2}}d_{gr} & \frac{1}{\sqrt{2}}d_{gg} \\ \hline
         \frac{1}{\sqrt{2}}d_{gg} & \frac{1}{\sqrt{2}}d_{gg} & & & & d_{rr} \\
         \frac{1}{\sqrt{2}}d_{gr} & \frac{-1}{\sqrt{2}}d_{gr} & & & d_{rg} & \\
         \frac{1}{\sqrt{2}}d_{rg} & \frac{-1}{\sqrt{2}}d_{rg} & & d_{gr} & & \\
         \frac{1}{\sqrt{2}}d_{rr} & \frac{1}{\sqrt{2}}d_{rr} & d_{gg} & & & \\
    \end{array}
    \right) \ ,
\end{equation}
where $d_{\alpha\beta} = \delta_{a\alpha}\delta_{b\beta}$. Importantly, $U_{ab}$ is not block-off diagonal (which is the case, for instance, of SU(2)): this implies that acting with the parallel transporter on a link in the fundamental representation $\tau$ can still result in the same irrep. 
In Fig.~\ref{fig:semilinks}(e), the arrows depict non-vanishing CG coefficients. Since for $D_3$ we have $3$ fluxes (between different representations), we need to formulate three rishon modes, one for each flux.
By reducing the representation basis to the left and right semilink basis, $\{ \ket{0}, \ket{p}, \ket{r}, \ket{g} \}$, one can check that the following rishon operators satisfy Eq.~\eqref{eq:rishon_decomposition} in the decomposition of the parallel transporter of Eq.~\eqref{eq:par_tranD3}:
\begin{align}
    \zeta_a^{(1)} = 2^{-\frac{1}{4}} \left(
    \begin{array}{c|c|cc}
         0 & & \delta_{ar} & \delta_{ag}  \\ \hline
         & 0 & 0 & 0  \\ \hline
         \delta_{ag} & 0 & 0 &  \\
         \delta_{ar} & 0 &   & 0
    \end{array}
    \right) \ , \hspace{5mm}
    \zeta_a^{(2)} = 2^{-\frac{1}{4}} \left(
    \begin{array}{c|c|cc}
         0 & & 0 & 0 \\ \hline
         & 0 & \delta_{ar} & -\delta_{ag}  \\ \hline
         0 & \delta_{ag} & 0 &  \\
         0 & -\delta_{ar} &   & 0
    \end{array}
    \right) \ , \hspace{5mm}
    \zeta_a^{(3)} = \left(
    \begin{array}{c|c|cc}
         0 & & 0 & 0 \\ \hline
         & 0 & 0 & 0  \\ \hline
         0 & 0 & 0 & \delta_{ar} \\
         0 & 0 & \delta_{ag}  & 0
    \end{array}
    \right) \ . \nonumber
\label{eq:rishons_D3}
\end{align}
To construct the gauge-invariant dressed sites, we need to first describe how the gauge transformations act on the rishons.
It is sufficient to consider the transformations $\theta^{L(R)}_r$ and $\theta^{L(R)}_s$, with respect to the group elements $r$ and $s$, since all other transformations of $D_N$ can be written as products of their powers, 
\begin{equation}
\hspace{-0.8cm}
\begin{aligned}\label{eq:r_and_s_transformations}
    &\theta^L_r = \left( 
    \begin{array}{c|c|cccc}
         1 & &  &  &  &  \\ \hline
         & 1 &  &  &  &  \\ \hline
          &  & \omega & & &  \\
          & & & \omega &  & \\
          &  & &  & \omega^{-1} & \\
          &  &  & & & \omega^{-1} \\
    \end{array}
    \right) \ , \ \theta^L_s = \left( 
    \begin{array}{c|c|cccc}
         1 & &  &  &  &  \\ \hline
         & -1 &  &  &  &  \\ \hline
          &  & 0 & & 1 &  \\
          & & & 0 &  & 1 \\
          &  & 1 &  & 0 & \\
          &  &  & 1 & & 0 \\
    \end{array}
    \right) \ ,  \\
    &\theta^R_r = \left( 
    \begin{array}{c|c|cccc}
         1 & &  &  &  &  \\ \hline
         & 1 &  &  &  &  \\ \hline
          &  & \omega^{-1} & & &  \\
          & & & \omega &  & \\
          &  & &  & \omega^{-1} & \\
          &  &  & & & \omega \\
    \end{array}
    \right) \ , \ \theta^R_s = \left( 
    \begin{array}{c|c|cccc}
         1 & &  &  &  &  \\ \hline
         & -1 &  &  &  &  \\ \hline
          &  & 0 & 1 & &  \\
          & & 1 & 0 &  & \\
          &  & &  & 0 & 1 \\
          &  &  & & 1 & 0 \\
    \end{array}
    \right) \ ,
\end{aligned}
\end{equation}
where $\omega = \exp(2\pi i/3)$.
In the $4$-dimensional rishon basis $\{\ket{0}, \ket{p},\ket{r}, \ket{g} \}$, the transformations in Eq.~\eqref{eq:r_and_s_transformations} read
\begin{align}\label{eq:rish_tr}
    \Lambda_r^L = \left(
    \begin{array}{c|c|cc}
         1 & &  &  \\ \hline
         & 1 &  &   \\ \hline
          &  & \omega &  \\
          &  &   & \omega^{-1}
    \end{array}
    \right) \ , \
    \Lambda_r^R = \left(
    \begin{array}{c|c|cc}
         1 & &  &  \\ \hline
         & 1 &  &   \\ \hline
          &  & \omega^{-1} &  \\
          &  &   & \omega
    \end{array}
    \right) \ ,\ 
    \Lambda_s^L = \left(
    \begin{array}{c|c|cc}
         1 & &  &  \\ \hline
         & -1 &  &   \\ \hline
          &  & 0 & 1 \\
          &  & 1 & 0
    \end{array}
    \right) = \Lambda_s^R \ .
\end{align}

On the upper and lower dressed sites, the transformations read
\begin{align}
    \Lambda_{r/s}^{\rm up} = \Lambda_{r/s}^{R}\otimes\Lambda_{r/s}^{R}\otimes\Lambda_{r/s}^{L} \ , \notag\\
    \Lambda_{r/s}^{\rm down} = \Lambda_{r/s}^{R}\otimes\Lambda_{r/s}^{L}\otimes\Lambda_{r/s}^{L}  \ ,
\label{eq:trafo_semilinks}
\end{align}
while for the four corners, the transformation laws are 
\begin{align}
\Lambda_{r/s}^{\rm up, left} &= \Lambda_{r/s}^{R}\otimes\Lambda_{r/s}^{L} \ ,\:\:\:\Lambda_{r/s}^{\rm up, right} = \Lambda_{r/s}^{R}\otimes\Lambda_{r/s}^{R} \ ,\notag\\
\Lambda_{r/s}^{\rm down, left} &= \Lambda_{r/s}^{L}\otimes\Lambda_{r/s}^{L} \ ,\:\:\:\Lambda_{r/s}^{\rm down, right} = \Lambda_{r/s}^{R}\otimes\Lambda_{r/s}^{L} \ .
\label{eq:trafo_semilinks_edges}
\end{align}

The local gauge-invariant states in the neutral sector are constructed by projecting the semilinks attached to a vertex on the eigenspace of the gauge transformation operators from Eq.~\eqref{eq:trafo_semilinks} and Eq.~\eqref{eq:trafo_semilinks_edges} with eigenvalue $1$:
\begin{align}
    \Lambda_{r}\Lambda_{ s}\ket{\psi_{\rm phys}} = \ket{\psi_{\rm phys}} \ .
\end{align}
The order of projection with respect to the group elements is not important, as long as the physical space is the common eigenspace with real eigenvalues.

In the presence of charges, physical states on the vertices are constructed by taking the tensor product of the attached semilinks and the four-dimensional Hilbert space of the charge (charges being in the fundamental representation), 
\begin{align}
\Lambda_{r}\theta_{r}^{\psi}\Lambda_{s}\theta_{s}^{\psi}\ket{\psi_{\rm phys}} = \ket{\psi_{\rm phys}} \ ,
\end{align}
where we defined the transformation laws for the matter to be the same as for the right rishons (see Eq.~\eqref{eq:rish_tr}):
\begin{align}
    \theta_{r}^{\psi} = \begin{pmatrix}
      1 & 0 & 0 & 0 \\
      0 & 1 & 0 & 0 \\
      0 & 0 & \omega^{-1} & 0\\
      0 & 0 & 0 & \omega
    \end{pmatrix},\:\:\:
  \theta^{\psi}_{s} = \begin{pmatrix}
      1 & 0 & 0 & 0 \\
      0 & -1 & 0 & 0 \\
      0 & 0 & 0 & 1\\
      0 & 0 & 1 & 0
    \end{pmatrix}.
\end{align}

In $D_3$, in the charge-neutral sector, the gauge-invariant basis for the dressed sites at any vertex within the ladder is 11-dimensional. On the upper leg, labeling the semilinks to the left of the vertex, below the vertex, and to the right of the vertex sequentially, the basis states can be written as
\begin{align}\label{eq:dressed_basis_app}
    &\ket{1} = \ket{0,0,0} \quad \ket{2} = \frac{\ket{r,0,r}+\ket{g,0,g}}{\sqrt{2}} \nonumber \\
    &\ket{3} = \frac{\ket{0,r,r}+\ket{0,g,g}}{\sqrt{2}} \quad \ket{4} = \frac{\ket{r,g,0}+\ket{g,r,0}}{\sqrt{2}} \nonumber \\
    &\ket{5} = \frac{\ket{r,r,g}+\ket{g,g,r}}{\sqrt{2}} \quad \ket{6}=\frac{\ket{r,p,r}-\ket{g,p,g}}{\sqrt{2}} \\
    &\ket{7}=\frac{\ket{p,r,r}-\ket{p,g,g}}{\sqrt{2}} \quad \ket{8}=\frac{\ket{r,g,p}-\ket{g,r,p}}{\sqrt{2}} \nonumber \\
    &\ket{9} = \ket{p,p,0} \quad \ket{10} = \ket{p,0,p}  \quad \ket{11} = \ket{0,p,p} \nonumber \ .
\end{align}
For the dressed sites on the lower leg, the gauge-invariant basis can be obtained by conjugating the colored central fields $r\leftrightarrow g$ in Eq.~\eqref{eq:dressed_basis_app}.

The procedure of projecting on the gauge invariant subspace on each vertex can also be used for $D_4$, which gives a local gauge-invariant dimension of $28$ for each dressed site. On the corners, the dressed local dimension is, for $D_3$, $5$ with static charges and $3$ without, while for $D_4$ it is $8$ and $5$, respectively.
\subsection{Hamiltonian in the rishon basis}\label{app:h_rishon}
We proceed with projecting the Hamiltonian on the gauge-invariant dressed basis constructed above. Each semilink contributes with half the electric field energy. 
Thus, the electric Hamiltonian $H_E= \frac{c \hbar}{a} g^2 \sum_{l \in {\rm links}} \sum_j \alpha_j \Pi^j_l $, which is diagonal in the representation basis, can be reformulated as 
\begin{align}
    H_E= \frac{c \hbar}{a} g^2 \sum_{sl \in {\rm semilinks}} \sum_j \frac{\alpha_j}{2} \Pi^j_{sl} \ .
\end{align}
By considering the semilinks attached to each vertex, it is possible to express the electric Hamiltonian in terms of dressed sites
\begin{equation}
    H_E = \sum_n H_{E,\textbf{n}}^{(1)} \ , 
\end{equation}
where $H_E^{(1)}$ is the electric Hamiltonian acting on the single dressed sites. In $D_3$, in the 11-dimensional dressed basis in Eq.~\eqref{eq:dressed_basis_app}, its diagonal entries are the following:
\begin{align}
&H_E^{(1)} = \frac{c\hbar}{a}\frac{g^2}{2} {\rm diag}(3\alpha_0, \ 2\alpha_{\tau}+\alpha_0,\ 2\alpha_{\tau}+\alpha_0,\ 2\alpha_{\tau}+\alpha_0,\notag\\
&\quad3\alpha_{\tau},\ 2\alpha_{\tau}+\alpha_p, \ 2\alpha_{\tau}+\alpha_p, \ 2\alpha_{\tau}+\alpha_p, \ 2\alpha_p+\alpha_0, \notag \\
&\quad2\alpha_p+\alpha_0, \ 2\alpha_p+\alpha_0) \ .
\end{align}
The coefficients $\alpha_{j_i}$, which for Lie groups are fixed by the quadratic Casimir, can be chosen with a higher degree of freedom for discrete groups. However, one can construct a group Laplacian also in this case~\cite{Mariani_PRD2023}, restricting the choice of irrep energies to a few discrete sets. In our numerical simulations, we set the values $\alpha_0 = 0, \ \alpha_{\tau}=1, \ \alpha_p=10$.

The plaquette operator can be projected on the gauge-invariant dressed basis by employing the rishon formulation of the group connection:
\begin{align} 
&\sum_{a,b,c,d}\mathfrak{R}(U^{ab}_{\textbf{n},\textbf{n}+\textbf{e}_{\rm x}}U^{bc}_{\textbf{n}+\textbf{e}_{\rm x},\textbf{n}+\textbf{e}_{\rm x}+\textbf{e}_{\rm y}}U^{\dagger c d}_{\textbf{n}+\textbf{e}_{\rm y},\textbf{n}+\textbf{e}_{\rm x}+\textbf{e}_{\rm y}} U^{\dagger d a }_{\textbf{n},\textbf{n}+\textbf{e}_{\rm y}})\notag\\ 
&= \bigg(\sum_i\zeta^{(i)}_{a;\textbf{n},\rm x}\zeta^{(i)\dagger}_{\bar{b};\textbf{n}+\textbf{e}_{\rm x},-\rm x}\bigg)\bigg(\sum_i\zeta^{(i)}_{b;\textbf{n}+\textbf{e}_{\rm x},\rm y}\zeta^{(i)\dagger}_{\bar{c};\textbf{n}+\textbf{e}_{\rm x}+\textbf{e}_{\rm y},-\rm y}\bigg)\notag\\
&\times\bigg(\sum_i\zeta^{(i)}_{\bar{c};\textbf{n}+\textbf{e}_{\rm x}+\textbf{e}_{\rm y},-\rm x}\zeta^{(i)\dagger}_{d;\textbf{n}+\textbf{e}_{\rm y},\rm x}\bigg)\bigg(\sum_i\zeta^{(i)}_{\bar{d};\textbf{n}+\textbf{e}_{\rm y},-\rm y}\zeta^{(i)\dagger}_{a;\textbf{n},\rm y}\bigg)\,.
\end{align}
By reordering and regrouping the rishons, we can write down the plaquette in the following form:
\begin{align}
    &\sum_{a,b,c,d}\mathfrak{R}(U^{ab}_{\textbf{n},\textbf{n}+\textbf{e}_{\rm x}}U^{bc}_{\textbf{n}+\textbf{e}_{\rm x},\textbf{n}+\textbf{e}_{\rm x}+\textbf{e}_{\rm y}}U^{\dagger c d}_{\textbf{n}+\textbf{e}_{\rm y},\textbf{n}+\textbf{e}_{\rm x}+\textbf{e}_{\rm y}} U^{\dagger d a }_{\textbf{n},\textbf{n}+\textbf{e}_{\rm y}})\notag\\ = 
    &\sum^{3}_{i,j,k,l = 1}C^{li}_{\textbf{n},\rm x,\rm y} C^{ij}_{\textbf{n}+\textbf{e}_{\rm x}, \rm y,-\rm x} C^{jk}_{\textbf{n}+\textbf{e}_{\rm x}+\textbf{e}_{\rm y},-\rm x,-\rm y} C^{kl}_{\textbf{n}+\textbf{e}_{\rm y},-\rm y,\rm x} \ ,
\end{align}
where we defined the gauge-invariant corner operators as
\begin{align}
    C^{ij}_{\textbf{n},\mu_1, \mu_2} \equiv \sum_{a}\zeta^i_{a;\textbf{n},\mu_1}\zeta^{j\dagger}_{a;\textbf{n},\mu_2} \ ,
\end{align}
with $a \rightarrow \bar{a}$ when $\mu = -\rm x, \ -\rm y$.
The corner operators act per construction on the semilinks attached to the vertex on which they are defined. Furthermore, since they are gauge-invariant per construction (summation over color indices), the projection of the gauge-invariant subspace is rather trivial. 
Therefore, we can write the magnetic part of the Hamiltonian in the dressed site as 
\begin{align}
    H_{B} = -\frac{1}{ag^2}\sum_{\textbf{n};i,j,k,l}C^{li}_{\textbf{n},\rm x,\rm y}C^{ij}_{\textbf{n}+\textbf{e}_{\rm x}, \rm y,-\rm x}
    C^{jk}_{\textbf{n}+\textbf{e}_{\rm x}+\textbf{e}_{\rm y},-\rm x,-\rm y}C^{kl}_{\textbf{n}+\textbf{e}_{\rm y},-\rm y,\rm x},
\end{align}
where the summation is performed over the sites  ($\textbf{n}$) and rishon indices ($i,j,k,l$).

The non-Abelian Gauss's law is already satisfied in the construction of the dressed basis in Eq.~\eqref{eq:dressed_basis_app}. However, the splitting $\ket{j,m_L,m_R}= \ket{j_1,m_L}\otimes\ket{j_2,m_R}\delta_{j_1,j_2}\delta_{j,j_1}$ introduces constraints between adjacent dressed sites, since the left and right semilink representation of each link must be the same. In the effective Hamiltonian implemented in the DMRG, we add the following Abelian selection rules: 
\begin{align}\label{eq:penalties}
    &(\hat{D}_{\textbf{n}}^R\hat{D}_{\textbf{n}+\textbf{e}_{\rm x}}^L-1) \ket{\psi_{\rm phys}} = 0 \ \forall \bf{n} \ , \nonumber \\
    &(\hat{D}_{\textbf{n}}^C\hat{D}_{\textbf{n}+\textbf{e}_{\rm y}}^{C'}-1) \ket{\psi_{\rm phys}} = 0 \ \forall \bf{n} \ ,
\end{align}
where the $\hat{D}$ operators are diagonal in the dressed basis space, with entries that depend on the irrep in the left, right, and central semilinks of the dressed site state. Reminding that, by construction, $\hat{D}^L$ operators act on the right-most label in the dressed basis in Eq.~\eqref{eq:dressed_basis_app}, a possible choice for the $\hat{D}$ operators is
\begin{align}
    &\hat{D}^L_{\textbf{n}} = {\rm diag}(1,2,2,1,2,2,2,-1,1,-1,-1) \ , \nonumber \\ 
    &\hat{D}^R_{\textbf{n}} = {\rm diag}(1,\frac{1}{2},1,\frac{1}{2},\frac{1}{2},\frac{1}{2},-1,\frac{1}{2},-1,-1,1) \ ,           \\
    &\hat{D}^C_{\textbf{n}} = {\rm diag}(1,1,2,2,2,-1,2,2,-1,1,-1) \ , \nonumber \\
    &\hat{D}^{C'}_{\textbf{n}} = {\rm diag}(1,1,\frac{1}{2},\frac{1}{2},\frac{1}{2},-1,\frac{1}{2},\frac{1}{2},-1,1,-1) \ , \nonumber
\end{align}
where we assigned arbitrary values to the identity, parity, and fundamental representations, $(1,-1,2)$ respectively in $\hat{D}^L$ and $\hat{D}^C$, such that Eqs.~\eqref{eq:penalties} are satisfied only when the two semilinks are in the same irrep.

\subsection{Fermionic parity of the rishons}\label{app:parity}
One important consequence of the absence of center in the $D_{N_{\rm odd}}$ case is that the gauge group does not preserve local $\mathbb{Z}_2$ parity. The implication of this is that, when coupled to fermionic matter, the $D_{N_{\rm odd}}$ lattice gauge theory cannot be defermionized, i.e., it can not be mapped onto a local Hamiltonian without fermionic degrees of freedom independently of the dimensionality. 
Instead, in $D_{N_{\rm even}}$, the rishons are fermions and they can be combined with fermionic matter to form singlets, which behave as an effective bosonic particle.

Here, we illustrate this general property of dihedral groups on the example of $D_3$ and $D_4$---the latter having a $\mathbb{Z}_2$ center, while the former does not.
In the $D_{N_{\rm even}}$ case, the parity operator on a semilink is defined as
\begin{align}
    P = \left(
    \begin{array}{c|c|c|c}
         1 & & & \\ \hline
         & 1 & &  \\ \hline
         & & \ddots & \\ \hline
         & & & -\mathbb{1}_2
    \end{array}
    \right) \ .
\end{align}
$D_N$ only has one- and two-dimensional representations. The former ones carry even parity and the latter ones odd parity.

Similar to what we did above for $D_3$, from Fig.~\ref{fig:semilinks}(e) it is possible to count the number of different rishon modes, counting the number of arrows, to reconstruct the parallel transporter for $D_4$. A possible choice for the four rishon operators is
\begin{align}\label{eq:rishons_D4}
    \zeta_a^{(1)} &= 2^{-\frac{1}{4}} \left(
    \begin{array}{c|c|c|c|cc}
         0 & & & & \delta_{ar} & \delta_{ag}  \\ \hline
         & 0 & & & &  \\ \hline
         & & 0 & & &  \\ \hline
         & & & 0 & &  \\ \hline
         \delta_{ag} & & & & 0 & \\
         \delta_{ar} & & & & & 0
    \end{array}
    \right) \ , \hspace{10mm}
    \zeta_a^{(2)} = 2^{-\frac{1}{4}} \left(
    \begin{array}{c|c|c|c|cc}
         0 & & & &  & \\ \hline
         & 0 & & & \delta_{ar} & -\delta_{ag}  \\ \hline
         &  & 0 & & & \\ \hline
         &  & & 0 & &  \\ \hline
         & \delta_{ag} & & & 0 \\
         & -\delta_{ar} & & & & 0
    \end{array}
    \right) \ , \nonumber \\
    \zeta_a^{(3)} &= 2^{-\frac{1}{4}} \left(
    \begin{array}{c|c|c|c|cc}
         0 & & & &  & \\ \hline
         & 0 & & &  &   \\ \hline
         &  & 0 & & \delta_{ag} & -\delta_{ar} \\ \hline
         &  & & 0 & &  \\ \hline
         & &\delta_{ar}  & & 0 \\
         & &-\delta_{ag}  & & & 0
    \end{array}
    \right) \ , \hspace{5mm}
    \zeta_a^{(4)} = 2^{-\frac{1}{4}} \left(
    \begin{array}{c|c|c|c|cc}
         0 & & & &  & \\ \hline
         & 0 & & &  &   \\ \hline
         &  & 0 & & & \\ \hline
         &  & & 0 & \delta_{ag} & \delta_{ar}  \\ \hline
         & & & \delta_{ar} & 0 \\
         & & & \delta_{ag} & & 0
    \end{array}
    \right) \ .
\end{align}
The rishons formulated in Eq.~\eqref{eq:rishons_D4} for the $D_4$ theory anti-commute with $P$ on the same semilink and commute on different semilinks. Therefore, one can define fermionic rishon operators by attaching a Jordan--Wigner-like parity string to the local rishon operators,
\begin{align}
    \zeta^{(i)}_{F,a;\textbf{n},\mu} = \bigg(\bigotimes_{({\bf n}', \mu')\in \rm path}P_{{\bf n}', \mu'} \bigg)\otimes \tilde{\zeta}^{(i)}_{a;\textbf{n},\mu}\otimes \mathbb{1} \ ,
\label{eq:fermionic_rishons}
\end{align}
where we denoted
\begin{align}
    \tilde{\zeta}^{(i)}_{a;\textbf{n},\mu} = \begin{cases}
    \zeta^{(i)}_{a;\textbf{n},\mu}P_{\textbf{n},\mu}\,, \quad \text{on left semilinks,} \\
    \zeta^{(i)}_{a;\textbf{n},\mu}\,, \quad \text{on right semilinks.}
    \end{cases}\\
\end{align}
In Eq.~\eqref{eq:fermionic_rishons}, the path, as illustrated in Fig.~\ref{fig:semilinks}, is a particular choice of ordering the semilinks on the lattice. 
The rishons defined as in Eq.~\eqref{eq:fermionic_rishons} anti-commute on different semilinks,
\begin{align}
    \{\tilde{\zeta}^{(i)}_{F,a;{\bf n}, \mu},\tilde{\zeta}^{(i)}_{F,a;{\bf n}', \mu'}\} = 0 \:\: \text{for } ({\bf n}, \mu)\neq ({\bf n}', \mu')\,,
\end{align}
and therefore are indeed of fermionic nature. The local rishon operators can simply be replaced by the fermionic ones in the expression for the link operator (Eq.~\eqref{eq:rishon_decomposition}):
\begin{align}\label{eq:rishon_F_decomposition}
    U_{ab;\textbf{n},\textbf{n}+\textbf{e}_{\rm \mu}} = \sum_i \zeta_{F,a;\textbf{n},\rm \mu}^{(i)} \zeta_{F,\bar{b};\textbf{n}+\textbf{e}_{\rm \mu},\rm -\mu}^{(i)\dagger} \ .
\end{align}
In the presence of dynamical matter fermions, the hopping term in the rishon formulation becomes
\begin{align}
    \psi_{a,\textbf{n}}^{\dagger}U_{ab;\textbf{n},\textbf{n}+\textbf{e}_{\rm \mu}}\psi_{b,\textbf{n}+\textbf{e}_{\rm \mu}} + \mathrm{h.c.}= \sum_i\psi^{\dagger}_{a,\textbf{n}}\zeta^{(i)}_{F,a;\textbf{n},\rm \mu}\zeta^{(i)\dagger}_{F,\bar b;\textbf{n}+\textbf{e}_{\rm \mu},\rm -\mu}\psi_{b,\textbf{n}+\textbf{e}_{\rm \mu}} + \mathrm{H.c.}
\end{align}
The right-hand side of the expression above is a product of operators that commute with the fermionic parity on each dressed site and, therefore, can be defermionized.

In the case of the $D_{N_{\rm odd}}$ gauge theory, the screening fusion rule prevents us from assigning odd parity to the fundamental representation. Regardless of whether we assign even or odd parity to the anti-fundamental representation $\taf$, the product on the left-hand side of $j\otimes \tau \ni j$ is of even parity. Then, the only possibility for the fundamental representation is to be of even parity, rendering some rishons in the decomposition of Eq.~\eqref{eq:rishon_decomposition} to be bosonic operators. Therefore, no full defermionization as in the $D_4$ (or, in general, the $D_{N_{\rm even}}$) case is possible.

\section{Weak confinement in $\mathbb{Z}_2$ ladder}\label{app:z2}
To show that the quasi 1d geometry can induce a weak confinement even in the weak-coupling regime, we analyze the case of a $\mathbb{Z}_2$ LGT ladder (see also Refs.~\cite{Nyhegn2021,Pradhan_PRB2024}).
The full Hamiltonian reads
\begin{equation}
H_{\mathbf{Z}_2}= -g^2 \sum_l\sum_{s=\uparrow,\downarrow,r} Z_{l,s} -\frac{1}{g^2}\sum_l X_{l,\downarrow}X_{l+1,r}X_{l,\uparrow}X_{l,r} \ ,
\end{equation}
where $l$ labels the plaquettes, $\uparrow/\downarrow$ the upper/lower legs, and $r$ the left rung. $Z$ and $X$ are Pauli operators.
Thanks to the presence of gauge symmetries, we can work in the dual lattice and define new Pauli operators $Z_l=Z_{l,\uparrow}$. Assuming that the background field is everywhere $Z_{\rm bg}=+1$, gauge symmetry imposes that $Z_{l,r}=Z_{l-1,\uparrow}Z_{l,\uparrow}=Z_{l-1}Z_{l}$. 
To write $Z_{l,\downarrow}$ as a function of $Z_l$, we have to choose whether we have charges on the corners of the ladder or not.
If not, physical states are closed electric loops, meaning that $Z_{l,\downarrow}=Z_l$. With charges, instead, they are open-ended strings pinned at the ladder corners, meaning that $Z_{l,\downarrow}=-Z_l$.
Finally, the plaquette operator flips the value of the electric field, acting in the dual lattice as a Pauli $X_l$ operator.
The Hamiltonian thus becomes
\begin{align}
    \tilde{H}_{\mathbf{Z}_2} = &- g^2 \sum_{l=1}^{R-1} Z_l Z_{l+1}- 2g^2 2(1-q)\sum_{l=1}^{R-1} Z_l 
     +g^2(1-q)(Z_1+Z_R)  - \frac{1}{g^2}\sum_l X_l \ , 
\end{align}
where $q=0$ ($q=1$) indicates the absence (presence) of static charges at the corners.
$R$ is the number of plaquettes in the ladder and becomes the number of spin variables in the dual Hamiltonian.

In the weak-coupling regime, the ground-state energy can be computed in perturbation theory over the polarized state $\otimes_l \ket{+}_l$.
Straightforward calculations lead to
\begin{align}
    E_{q=0} &= -\frac{R}{g^2} - g^6 \frac{9R+19}{4} + o(g^6) \ ,\\ 
    E_{q=1} &= -\frac{R}{g^2} - g^6 \frac{R+3}{4} + o(g^6) \ .
\end{align}
The energy difference is therefore $E_{q}-E_{q=0}=2g^6(R+2)$, which is proportional to the separation of the two charges $R$, indicating some weak confinement even in the magnetic regime.
This contrasts the full 2d scenario, where the $\mathbb{Z}_2$ LGTs (and more broadly $\mathbb{Z}_N$) undergo a deconfinement phase transition at a finite coupling strength $g_c$ and the string tension vanishes exponentially for $g<g_c$~\cite{Fradkin_book}. The behavior at weak coupling on a ladder, therefore, is of geometric origin.
The fact that dihedral groups, see main text, display exactly the same $g^6$ scaling of the string tension at weak coupling suggests this is also a perturbative effect coming from geometric confinement.

\section{Comparison between SU(N) and $D_3$}\label{app:comparison}
\begin{figure}
    \centering
    \includegraphics[width=\linewidth]{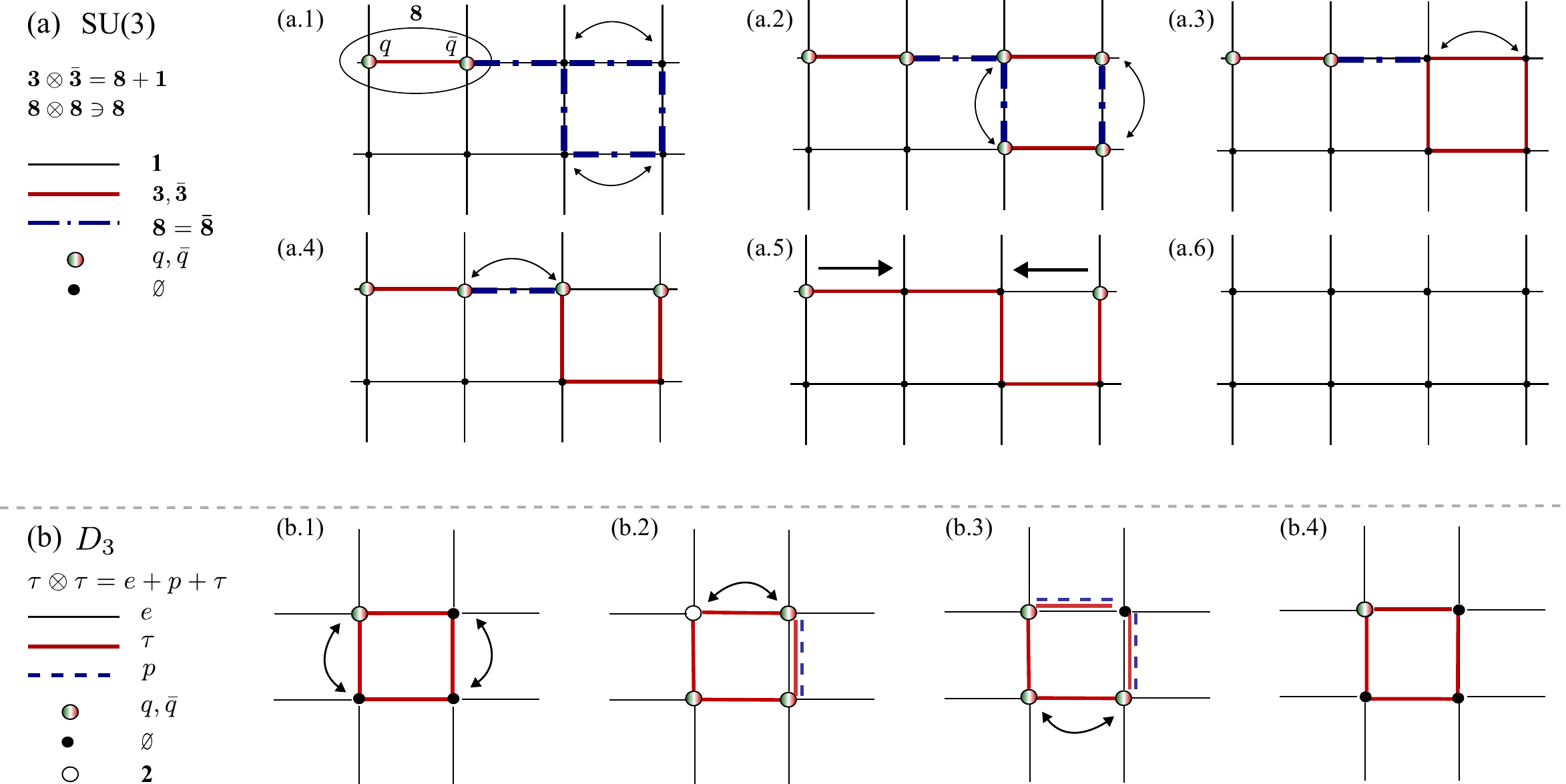}
    \caption{Gluelump stability with dynamical staggered fermions in the fundamental irreducible representation.
    In SU(3) (a), the gluelump is associated with a (bosonic) composite particle composed of a quark-antiquark pair $q\bar{q}$ in the adjoint representation $\bf{8}$. Since the bosonic number is not conserved, the gluelumps can be annihilated by a series of particle--antiparticle nucleations and annihilations (arrows).
    Conversely, in $D_3$ (b), vacuum fluctuations can not annihilate a gluelump, as its fermionic parity is preserved by the analogue processes.
    For both groups, we highlight the relevant fusion rules. Black circles on the sites correspond to the local vacuum following the staggered fermion prescription, i.e., empty for particles and fully occupied for antiparticles.
    }
    \label{fig:D3vsSU3}
\end{figure}
In this appendix, we elaborate further the comparison between SU($N$) and $D_N$ gluelumps. To remark the difference between $D_N$ with $N$ odd and SU($N$), it is instructive to look also at $D_N$ with $N$ even. Indeed, in $D_{N_{\rm even}}$ models, with $N_{\rm even}\ge 6$, there are one or more 2-dimensional irreducible representations that are non-faithful. These irreps play the same role as the adjoint representations in SU(N), since the center is mapped into the identity. As a consequence, gluelumps are legit gauge invariant states also in $D_{N_{\rm even}}$, with $N_{\rm even}\ge 6$ and the matter charge in non-faithful representations. The phenomenology of these theories is, then, closer to SU($N$) than to $D_{N_{\rm odd}}$. 

The possibility of defining gluelumps with fermionic matter in the fundamental representation of the group has important consequences on the stability of the gluelumps when the gauge theory is coupled to dynamical matter, highlighting the differences between $D_{N_{\rm odd}}$ and SU($N$). In QCD, adjoint matter is associated with composite objects like quark--antiquark pairs. 
Thus, on the lattice, SU($N$) gluelumps can be obtained by nucleating a quark--antiquark pair from the vacuum, moving either component \new{twice} around an arbitrary closed path, and merging it back to the partner particle. 
The reverse process would annihilate them; see Fig.~\ref{fig:D3vsSU3}(a1-a6) for a detailed breakdown of such dynamics on a lattice, based on a staggered fermion Kogut-Susskind construction of the SU(3) lattice model.
This is not the case, e.g., for the $D_3$ gauge theory with staggered fermions in the fundamental representation $\tau$. Here, the charge components of the gluelumps are subject to a global U(1) symmetry, protecting them from being absorbed in the vacuum, thus making the gluelumps stable particles, see Fig.~\ref{fig:D3vsSU3}(b1-b4).
The conclusion can be reached through fermionic parity conservation: the charge component of an SU(3) gluelump is bosonic. Thus, it has the same parity as the vacuum and the two can be connected by Hamiltonian processes. In $D_3$, instead, the matter component is fermionic and requires another charge to be annihilated into the vacuum.

We can now turn to non-Abelian discrete subgroups of SU(2) and SU(3), asking whether they allow for the construction of gluelumps with charge in the fundamental representation. In the case of SU(2), all crystal-like subgroups have the same non-trivial center as SU(2), and so single fundamental charges cannot be screened by gauge fields. Discrete subgroups of SU(3), instead, present a more complex picture. Some subgroups studied in the context of quantum simulation of LGTs, like $\Sigma(36\times3)$, $\Sigma(72\times3)$, $\Sigma(216\times3)$, and $\Sigma(360\times3)$, as well as $\Delta(27)$ and $\Delta(54)$, have a non-trivial center. On the contrary, $\Sigma(60)$ and $\Sigma(168)$~\cite{Grimus_JoPA2010} have a trivial center, allowing, in principle, string breaking by gluelump formation.

FSU3D3

\end{document}